\documentclass[aps,floatfix,preprint,nofootinbib,superscriptaddress,natbib]{revtex4}

\usepackage{graphicx,float}
\usepackage{epsfig}		
\usepackage{dcolumn}
\usepackage{bm}
\usepackage{subfigure}
\usepackage[all]{xy}
\usepackage{amsmath,upgreek}
\usepackage{amssymb}

\usepackage{pdfpages}
\usepackage{color}
\usepackage{graphicx,epstopdf}
\usepackage[colorlinks=true,
 linkcolor=red,
 urlcolor=purple,
 citecolor=blue,hyperindex]{hyperref}
\def\0{\mbox{\tiny $0$}}
\def\1{\mbox{\tiny $1$}}
\def\2{\mbox{\tiny $2$}}
\def\3{\mbox{\tiny $3$}}
\def\4{\mbox{\tiny $4$}}
\def\5{\mbox{\tiny $5$}}
\def\6{\mbox{\tiny $6$}}
\def\7{\mbox{\tiny $7$}}
\def\8{\mbox{\tiny $8$}}
\def\9{\mbox{\tiny $9$}}

\def\f14{\mbox{\tiny $\frac{1}{4}$}}

\begin{document}

\title{Geometrical structure of the Wigner flow information quantifiers and hyperbolic stability in the phase-space framework}

\author{Alex E. Bernardini}
\email{alexeb@ufscar.br}
\altaffiliation[On leave of absence from]{~Departamento de F\'{\i}sica, Universidade Federal de S\~ao Carlos, PO Box 676, 13565-905, S\~ao Carlos, SP, Brasil.}
\affiliation{Departamento de F\'isica e Astronomia, Faculdade de Ci\^{e}ncias da
Universidade do Porto, Rua do Campo Alegre 687, 4169-007, Porto, Portugal.}
\date{\today}

\begin{abstract}
Quantifiers of stationarity, classicality, purity and vorticity are derived from phase-space differential geometrical structures within the Weyl-Wigner framework, after which they are related to the hyperbolic stability of classical and quantum-modified Hamiltonian (non-linear) equations of motion. By examining the equilibrium regime produced by such an autonomous system of ordinary differential equations, a correspondence between Wigner flow properties and hyperbolic stability boundaries in the phase-space is identified. Explicit analytical expressions for equilibrium-stability parameters are obtained for quantum Gaussian ensembles, wherein information quantifiers driven by Wigner currents are identified. Illustrated by an application to a Harper-like system, the results provide a self-contained analysis for identifying the influence of quantum fluctuations associated to the emergence of phase-space vorticity in order to quantify equilibrium and stability properties of Hamiltonian non-linear dynamics.
\end{abstract}

\maketitle

\section{Introduction}

The phase-space formulation of quantum mechanics provided by the Weyl-Wigner (WW) framework \cite{Wigner,Case} is governed by a consistent quantization procedure that maps analytic functions on the classical phase space to operators in the Hilbert space. This mapping, facilitated by the Weyl transform \cite{Weyl}, establishes a bridge between operator methods and path integral techniques \cite{Abr65,Sch81,Par88}. Beyond its phenomenological applications in the analysis of information flow dynamics in fields such as quantum optics, plasma physics, and nuclear physics \cite{Sch,Lvovsky2009}, and in addressing scattering and decoherence issues in semiconductor systems \cite{Zachos2005,Jung09}, the WW framework also provides fundamental tools for deepening the understanding of standard quantum mechanics \cite{Catarina,Bernardini13B,PhysicaA}, particularly due to its extensive connections with quantum information measures.
The WW framework extends to an enriched interpretation of wave-function collapse \cite{Zurek02,Bernardini13A,Leal2019}, a generalization of the correspondence between uncertainty relations and quantum observables \cite{Stein,Bernardini13C,Bernardini13E}, and elucidates connections between quantum mechanics and classical statistical mechanics, notably within the realms of quantum chaos \cite{Chaos} and quantum cosmology \cite{JCAP18}. By circumventing the operator formalism, this approach renders quantum mechanics in a manner closely analogous to classical Hamiltonian mechanics, aiding in the identification of classical-quantum boundaries in physical systems.
Likewise, without compromising the predictive capacity of quantum mechanics, the WW formalism can also be interpreted as a fluid-like description \cite{Ballentine,Steuernagel3,NossoPaper,NossoPaper19,Meu2018} of phase-space information flow, in terms of continuity equations that bridge classical and quantum regimes.

In such a context, some interesting quantifiable issues can also be identified for non-linear systems driven $1$-dim phase-space Hamiltonians cast as $H^{W}(q,\,p) = K(p) + V(q)$, when they are not encompassed by Schr\"odinger position and momentum representation pictures, i.e. if $K(p)$ replaces either linear or quadratic momentum contributions by either higher order or even non-polynomial functions.
The quantified flux of information \cite{NossoPaper,NossoPaper19,JCAP18,Meu2018,Bernardini2020-02}, in such a context, refers to differential operators providing a quantitative view of classicality, stationarity, purity and equilibrium/stability properties, for which exact corrections due to quantum fluctuations over a classical background will be found in terms of the geometrical structure of the Wigner flow. 
In addition, from the same geometrical structure, non-linear entropy and vorticity quantifiers can also be investigated in a similar context involving Wigner functions and Wigner currents.

Besides revisiting the computation of quantifiers of stationarity, classicality and purity, now extended to the above mentioned generalized Hamiltonian systems, with $H^{W}(q,\,p) = K(p) + V(q)$, the main issue discussed in this letter is related to the phase-space (velocity field) vorticity quantifier, which is shown to be related to the hyperbolic stability parameters, and essential in identifying classical and quantum-modified regimes driven by $H^{W}(q,\,p)$.

The generalized results of this work are specialized to the Harper-like systems which, besides working as a feasible platform for testing the classical-quantum boundaries, resume a broad class of non-linear Hamiltonian systems which can be addressed by the WW phase-space formalism.
As a a feasible platform for testing the identified classical-quantum boundaries, Harper-like Hamiltonians \cite{Harper01,Harper01BB,Harper01CC} are considered. 
The Harper framework describes the lattice pattern of electrons coupled to magnetic fields in planar configurations, which work, for instance, as a subjacent framework \cite{Nat010,Nat020} for parameterizing fractal structures related to the Hofstadter spectral decomposition. More recently, it has supported the investigation of quantum mechanical topological phases in the context of {\em engendering} ultra-cold atom platforms for emulating synthetic gauge fields and topological structures for neutral atoms \cite{001,002}, as well as the understanding of subtle features of the quantum Hall effect \cite{Prange}.
Due to non-linear contributions arising from the momentum coordinate, $p$, the Harper-like kinetic Hamiltonian contribution, $K(p)$, is displayed in the sinusoidal form of $\cos(p/p_0)$, which introduces manipulable non-linear features.

The outline of the manuscript is then as follows.
The differential geometrical structure of the Wigner flow quantifiers, their straightforward relations with the associated dynamical system Wigner flow quantifiers related to stationarity, classicality, purity and vorticity are described in Sec. II.
Part of this content revisits some of our previous results \cite{NossoPaper,NossoPaper19,JCAP18,Meu2018,Bernardini2020,Bernardini2020-02} from which quantum information fluxes are described by means of continuity equations identified and expressed in terms of Wigner functions and Wigner currents.
In Sec. III, after re-obtaining the Wigner framework for non-linear Hamiltonian systems, the correspondence of the same dynamical properties with hyperbolic stability parameters is provided.
In the context of discussing the hyperbolic stability affected by quantum fluctuations, explicit and generic analytical expressions are obtained for Gaussian ensembles driven by the corresponding Winger currents. From Wigner flow patterns obtained for a Harper-like Hamiltonian system, a summary of stability conditions is then obtained.
Our conclusions and an outlook for the next steps are presented in Sec. IV.

\section{Wigner function and Wigner flow quantifiers}

The formulation of the phase-space quantum mechanics according to the Weyl-Wigner \cite{Wigner,Case} framework encompasses all the observable content obtained from the Schr\"odinger picture through a convolutional representation of the wave function described by the Weyl transform.
From the Weyl transform of a generic quantum operator, $\hat{O}$, defined by
\small\begin{eqnarray}
O^W(q,\, p)\label{eqnWigner111}
&=& 2\hspace{-.1cm} \int^{+\infty}_{-\infty} \hspace{-.45cm}ds\,\exp{\left[+2\,i \,p\, s/\hbar\right]}\,\langle q - s | \hat{O} | q + s \rangle = 2\hspace{-.1cm} \int^{+\infty}_{-\infty} \hspace{-.45cm} dr \,\exp{\left[-2\, i \,q\, r/\hbar\right]}\,\langle p - r | \hat{O} | p + r\rangle,
\end{eqnarray}\normalsize
where $\hbar = h/2\pi$ is the reduced Planck constant, the inception of the Wigner function, $W(q, p)$, follows from the transform of a density matrix operator, $\hat{\rho} = |\psi \rangle \langle \psi |$, as
\begin{eqnarray}
 h^{-1} \hat{\rho} \to W(q,\, p) 
&=& (\pi\hbar)^{-1} 
\int^{+\infty}_{-\infty} \hspace{-.15cm}ds\,\exp{\left[2\, i \, p \,s/\hbar\right]}\,
\psi(q - s)\,\psi^{\ast}(q + s),\label{eqnWigner222}
\end{eqnarray}
with $\psi(q-s)\,\psi^{\ast}(q+s) = \langle q - s | \psi \rangle \langle \psi | q + s \rangle = \langle q - s |\hat{\rho}| q + s \rangle$, and $W(q,\, p)$ also interpreted as the Fourier transform of the off-diagonal terms of the associated density matrix.
Even noticing that it is supported by the Hilbert space properties of quantum mechanics, the coexistence of a positive-definite position and momentum probability distribution is suppressed by the overlap integral from Eq.~\eqref{eqnWigner222}. 
The Wigner function, $W(q,\, p)$, exhibits the properties of a real-valued {\em quasi}-probability distribution, since it can indeed assume local negative values.
Otherwise, considering its correspondence with the Schr\"odinger position and momentum representations of a quantum particle, which are respectively described by complex amplitude functions, $\psi(q)$ and $\varphi(p)$, the Weyl-Wigner \cite{Wigner} representation of quantum mechanics provides an alternative phase-space $q-p$ description of quantum states
through which the quantum analog of a classical phase-space probability distribution can hence be interpreted from $W(q,\, p)$\footnote{In fact, the Wigner function marginal distributions return position and momentum distributions upon integrations over the conjugated coordinate,
\begin{equation}
\vert \psi(q)\vert^2 = \int^{+\infty}_{-\infty} \hspace{-.35cm}dp\,W(q,\, p)
\qquad
\leftrightarrow
\qquad
\vert \varphi(p)\vert^2 = \int^{+\infty}_{-\infty} \hspace{-.35cm}dq\,W(q,\, p),
\end{equation}
with the Fourier transforms of the respective complex amplitude functions being identified by
\begin{equation}
 \varphi(p)=
(2\pi\hbar)^{-1/2}\int^{+\infty}_{-\infty} \hspace{-.35cm} dq\,\exp{\left[+i \, p \,q/\hbar\right]}\,
\psi(q),\quad
\mbox{and}
\quad \psi(q)=
(2\pi\hbar)^{-1/2}\int^{+\infty}_{-\infty} \hspace{-.35cm} dp\,\exp{\left[-i \, p \,q/\hbar\right]}\,
 \varphi(p).
\end{equation}}.
Moreover, informational features of quantum nature can be examined without affecting the predictive power of the theory supported by the symmetries of the Heisenberg-Weyl group of translations.
These features are straightforwardly obtained from Eqs.~\eqref{eqnWigner111} and \eqref{eqnWigner222}. That is the case of the expectation values of quantum observables described by $\hat{O}$, evaluated from an overlap integral over the phase-space coordinates, $q$ and $p$, as \cite{Wigner,Case}
\begin{equation}
 \langle O \rangle = 
\int^{+\infty}_{-\infty} \hspace{-.35cm}dp\int^{+\infty}_{-\infty} \hspace{-.35cm} {dq}\,\,W(q,\, p)\,{O^W}(q,\, p), \label{eqneqfive}
\end{equation}
which is identified with the trace of the product between $\hat{\rho}$ and $\hat{O}$, as $Tr_{\{q,p\}}\left[\hat{\rho}\hat{O}\right]$ and from which one notices that $Tr_{\{q,p\}}[\hat{\rho}]=1$ -- the normalization condition consistent with a probability distribution interpretation.

The time-evolving properties of the Wigner function, $W(q,\,p) \to W(q,\,p;\,t)$, can also be mapped into a fluid-analog framework \cite{Steuernagel3} strictly connected to the Hamiltonian dynamics.
In particular, the flow field related to the Wigner function evolution is parameterized by the so-called Wigner current \cite{Steuernagel3,NossoPaper,NossoPaper19,Meu2018}, $\mbox{\boldmath${J}$}(q,\,p;\,t)$, vectorially decomposed into the phase-space coordinate directions, $\hat{q}$ and $\hat{p}$, as $\mbox{\boldmath${J}$} = J_q\,\hat{q} + J_p\,\hat{p}$, through the established continuity equation \cite{Case,Ballentine,Steuernagel3,NossoPaper,NossoPaper19,Meu2018},
\begin{equation}
{\partial_t W} + {\partial_q J_q}+{\partial_p J_p} =0,
\label{eqnalexquaz51}
\end{equation}
where a shortened notation for partial derivatives, $\partial_a \equiv \partial/\partial a$, is assumed.
For a generic Schr\"odinger-like Hamiltonian operator, ${H}(\hat{Q},\,\hat{P})$,
convoluted by the Weyl transform so as to return $H^{W}(q,\, p)$, with
\begin{equation}
H^{W}(q,\, p) = \frac{{p}^2}{2m} + V(q),
\end{equation}
one has \cite{Case,Ballentine,Steuernagel3,NossoPaper,NossoPaper19}
\begin{equation}
J_q(q,\,p;\,t)= \frac{p}{m}\,W(q,\,p;\,t), \label{eqnalexquaz500BB}
\end{equation}
and
\begin{equation}
J_p(q,\,p;\,t) = -\sum_{\eta=0}^{\infty} \left(\frac{i\,\hbar}{2}\right)^{2\eta}\frac{1}{(2\eta+1)!} \, \left[\partial_q^{2\eta+1}V(q)\right]\,\partial_p ^{2\eta}W(q,\,p;\,t),
\label{eqnalexquaz500}
\end{equation}
with $\partial^s_a \equiv (\partial/\partial a)^s$.

From Eq.~\eqref{eqnalexquaz500}, one notices that the quantum corrections which distort classical trajectories are due to the series expansion contributions from $\eta \geq 1$. 
Of course, the suppression of $\eta \geq 1$ contributions leads to the classical analog Hamiltonian description of probability distribution dynamics described by the phase-space current components in the classical limit,
\begin{equation}
J^{\mathcal{C}}_q(q,\,p;\,t)= +({\partial_p H^{W}})\,W(q,\,p;\,t), \label{eqnalexquaz500BB2}
\end{equation}
\begin{equation}
J^{\mathcal{C}}_p(q,\,p;\,t) = -({\partial_q H^{W}})\,W(q,\,p;\,t),
\label{eqnalexquaz500CC2}
\end{equation}
which, once substituted into Eq.~\eqref{eqnalexquaz51}, returns the so-called Liouville equation, where the phase-space velocity is identified by $\mathbf{v}_{\xi(\mathcal{C})} = \dot{\mbox{\boldmath $\xi$}} = (\dot{q},\,\dot{p})\equiv ({\partial_p H^{W}},\,-{\partial_q H^{W}})$, for a classical path $\mathcal{C}$, with $\mbox{\boldmath $\nabla$}_{\xi}\cdot \mathbf{v}_{\xi(\mathcal{C})}= \partial_q \dot{q} + \partial_p\dot{p} = 0$, where {\em dots} correspond to time derivatives, $d/dt$.

Considering that quantum mechanics deals with probabilities, while classical mechanics deals with phase-space trajectories, the Wigner continuity equation and the above fluid-analog correspondence with the classical mechanics are the drivers for examining the connection between quantum and classical dynamics.
For a quantum current parameterized by $\mbox{\boldmath${J}$} = \mathbf{w}\,W$, where a Wigner phase-space velocity, $\mathbf{w}$, is identified as the quantum analog of $\mathbf{v}_{\xi(\mathcal{C})}$, a quantifiable divergent behavior can be computed from \cite{Steuernagel3}
\begin{equation}
\mbox{\boldmath $\nabla$}_{\xi} \cdot \mathbf{w} = \frac{W\, \mbox{\boldmath $\nabla$}_{\xi}\cdot \mbox{\boldmath${J}$} - \mbox{\boldmath${J}$}\cdot\mbox{\boldmath $\nabla$}_{\xi}W}{W^2}.
\label{eqnzeqnz59}
\end{equation}

In this context, the stationary behavior and the classical limit of a quantum system described by the Wigner function, $W \equiv W(q,\,p;\,t)$, are both quantified through Eqs.~\eqref{eqnalexquaz51} and \eqref{eqnzeqnz59}, respectively by $\mbox{\boldmath $\nabla$}_{\xi} \cdot \mbox{\boldmath${J}$} = 0$ and $\mbox{\boldmath $\nabla$}_{\xi} \cdot \mathbf{w} = 0$.
Likewise, when the Wigner currents are cast in the form of $\mbox{\boldmath${J}$} = \mathbf{w}\,W$, several complementary topological properties of the Wigner flow can be accessed by divergent and rotational operations, $\mbox{\boldmath $\nabla$}_{\xi} \cdot$ and $\mbox{\boldmath $\nabla$}_{\xi} \times$ (cf. Appendix I), under both differential and integral perspectives, as it shall be verified in the following.

\subsubsection{Stationarity} The differential form of the stationarity condition is, of course, obtained from Eq.~(\ref{eqnalexquaz51}), by setting $\mbox{\boldmath $\nabla$}_{\xi}\cdot \mbox{\boldmath${J}$} = 0$, as to return
${\partial_t W} = 0$.
In this context, the corresponding integrated probability,
\begin{equation}
\sigma_{(\mathcal{C})} =\int_{V_{_{\mathcal{C}}}}dV\,{W},
\label{eqnquaz6022}
\end{equation}
can be evaluated from Eq.~\eqref{eqnquaz692} for $\beta = 1$, which results into
\begin{eqnarray}
 \frac{D}{Dt}\sigma_{(\mathcal{C})}
 &=& - \oint_{\mathcal{C}}d\ell\,\mbox{\boldmath${J}$}\cdot \mathbf{n}.\label{eqnquaz69222}
\end{eqnarray}
Given that the Green's theorem sets 
\begin{eqnarray}
\int_{V_{\mathcal{C}}}dV \,\mbox{\boldmath $\nabla$}_{\xi}\cdot\mbox{\boldmath${J}$} 
 &=& \oint_{\mathcal{C}}d\ell\,\mbox{\boldmath${J}$}\cdot \mathbf{n},\label{eqnquaz}
\end{eqnarray}
for $\mbox{\boldmath $\nabla$}_{\xi} \cdot \mbox{\boldmath${J}$} = 0$, the above result reflects the phase-space probability conservation for stationary states, 
\begin{eqnarray}
 \frac{D}{Dt}\sigma_{(\mathcal{C})}=0,
\end{eqnarray}
in a result that can be generalized to any closed path, $\mathcal{A}$, replacing $\mathcal{C}$.

\subsubsection{Classicality} Following an analogous procedure, for $\beta = 0$, Eq.~\eqref{eqnquaz692} returns
\begin{eqnarray}
\frac{D}{Dt}V_{\mathcal{C}} = \frac{D}{Dt} \int_{V_{\mathcal{C}}}dV
 &=& \int_{V_{\mathcal{C}}}dV \,\mbox{\boldmath $\nabla$}_{\xi}\cdot\mathbf{w} 
- \oint_{\mathcal{C}}d\ell\,\mathbf{w} \cdot \mathbf{n}= 0,\label{eqnquaz69222}
\end{eqnarray}
which reflects the $V_{\mathcal{C}}$ localized phase-space volume conservation, given that, in this case, the Green's theorem is more helpful in the form of
\begin{eqnarray}
\int_{V_{\mathcal{C}}}dV \,\mbox{\boldmath $\nabla$}_{\xi}\cdot\mathbf{w} 
 &=& \oint_{\mathcal{C}}d\ell\,\mathbf{w} \cdot \mathbf{n}.\label{eqnquaz69222B}
\end{eqnarray}

Considering that $\mathbf{v}_{\xi(\mathcal{C})} \cdot \mathbf{n} = 0$, Eq.~\eqref{eqnquaz69222B} accounts for averaged flux fluctuations of $\mathbf{w}$ through the one-dimensional classical surface, $\mathcal{C}$, which, of course, vanishes for $\mbox{\boldmath $\nabla$}_{\xi} \cdot \mathbf{w} = 0$, in a kind of soft integral version of the Liouvillian condition.
In particular, even for non-Liouvillian systems, with $\mbox{\boldmath $\nabla$}_{\xi} \cdot \mathbf{w} \neq 0$, the local quantum fluctuations with respect to the classical pattern could be averaged out by an eventual vanishing integrated contribution identified by
\begin{equation}\label{eqnquaz69222C}
\oint_{\mathcal{C}}d\ell\,\mathbf{w} \cdot \mathbf{n} = 0.
\end{equation}
Assuming the phase-space volume conservation, one can say that quantumness is identified by $\mbox{\boldmath $\nabla$}_{\xi} \cdot \mathbf{w} \neq 0$ even that classicality could be misinterpreted from Eq.~(\ref{eqnquaz69222C}). Eq.~(\ref{eqnquaz69222C}) should be read as classical and quantum mechanics constrained by the same statistical background.

\subsubsection{Purity and Entropy} Complementary statistical aspects deduced from the Wigner function definition can be noticed from the replacement of ${O^W}(q,\, p)$ by $W(q,\, p)$ into Eq.~\eqref{eqneqfive}.
It returns the quantum purity obtained as an analogous of the trace operation, $Tr_{\{q,p\}}[\hat{\rho}^2]$, in the form of
\begin{equation}
\mathcal{P} = Tr_{\{q,p\}}[\hat{\rho}^2] = 2\pi\hbar\int^{+\infty}_{-\infty}\hspace{-.35cm}dp\int^{+\infty}_{-\infty} \hspace{-.35cm} {dq}\,\,\,W(q,\, p)^2,
\label{eqneqpureza}
\end{equation}
which satisfies the pure state constraint, $Tr_{\{q,p\}}[\hat{\rho}^2] = Tr_{\{q,p\}}[\hat{\rho}] = 1$, with the $2\pi\hbar$ factor introduced due to dimensional reasons so as to quantitatively admit extensions from pure states to statistical mixtures through the Weyl-Wigner formalism \cite{Case,Hillery}.

From Eq.~\eqref{eqnquaz692}, for $\beta = 2$, one has
\begin{eqnarray}
\frac{1}{2\pi\hbar} \frac{D~}{Dt}\mathcal{P}_{(\mathcal{C})} &=& 
\frac{D~}{Dt}\left(\int_{V_{\mathcal{C}}}dV\, {W}^{2}\right)
= -\int_{V_{\mathcal{C}}}dV\,{W}^{2} \,\mbox{\boldmath $\nabla$}_{\xi}\cdot\mathbf{w} 
- \oint_{\mathcal{C}}d\ell\, {W}\,\left(\mbox{\boldmath${J}$}\cdot \mathbf{n}\right).\label{eqnquaz692DD}
\end{eqnarray}

A correspondent analysis could be applied to the informational content related to the Weyl-Wigner form of the dimensionless von Neumann entropy given by
\begin{equation}
\mathcal{S}_{{\tiny\mbox{vN}}}=\frac{\mathcal{S}^B_{{\tiny\mbox{vN}}}}{k_B} =-\int_{V}dV\,{W}\,\ln\vert {2\pi\hbar W}\vert,
\label{zzvquaz60}
\end{equation}
which, through an approximation valid for peaked distributions \cite{NossoPaper,NossoPaper19}, can be reduced to the the so-called linear entropy correlated to the quantum purity by
\begin{eqnarray}
\mathcal{S}_{{\tiny\mbox{vN}}} \approx 1 - \mathcal{P},
\label{zzvquaz63}
\end{eqnarray}
where, again, the $2\pi\hbar$ factor has been introduced due to dimensional reasons, and $k_B$ is the Boltzmann constant.
In this case, the related flux of information through the classical boundary, $\mathcal{C}$, is expressed by \cite{NossoPaper,NossoPaper19,Meu2018}
\begin{eqnarray}
\frac{D~}{Dt}\mathcal{S}_{{\tiny\mbox{vN}}(\mathcal{C})}
&=& \int_{V_{\mathcal{C}}}dV\,\ln|2\pi\hbar{W}| \,\mbox{\boldmath $\nabla$}_{\xi}\cdot\mbox{\boldmath${J}$} 
+\oint_{\mathcal{C}}d\ell\,\mbox{\boldmath${J}$}\cdot \mathbf{n}
\end{eqnarray}

Even a R\'enyi-like entropy, which generalizes the Shannon entropy, the Hartley entropy, the collision entropy and the minimal entropy \cite{Jizba}, can be re-interpreted in terms of Wigner functions as (cf. Appendix I)
\begin{eqnarray}
R_{\beta(\mathcal{C})} = \frac{1}{1-\beta}\ln
\left[
(2\pi\hbar)^{\beta-1}\int_{V_{\mathcal{C}}}dV\,W^\beta
\right],
\end{eqnarray}
where the logarithm basis is arbitrary\footnote{Omitting the dimensional regularization, it could also be written as \begin{equation}
R_{\beta(\mathcal{C})} = \frac{1}{1-\beta}\ln
\left[
\int_{V_{\mathcal{C}}}dV\,W^\beta
\right],
\end{equation}
with the factor ${(1-\beta})^{-1}\ln\left[(2\pi\hbar)^{\beta-1}\right]$ being absorbed by the $R_{\beta(\mathcal{C})}$ definition.}.

The index $\beta$ is here introduced to estimate an associated fractal dimension and, for $\beta > 1$, it can be
identified into the derivative structure from Eq.~\eqref{eqnquaz692} so as to parameterize the time dependence of the R\'enyi-like entropy in terms of \cite{NossoPaper,NossoPaper19}
\begin{eqnarray}
({1-\beta}){e^{[(1-\beta)R_{\beta(\mathcal{C})}]}}\frac{D}{Dt}{R_{\beta(\mathcal{C})}} &=& (2\pi\hbar)^{\beta-1}\frac{D~}{Dt}\left(\int_{V_{\mathcal{C}}}dV\, \mathcal{W}^{\beta}\right).
\end{eqnarray}

Just to summarize, for this set of preliminary results, two additional features can also be noticed.
Firstly, for stationary states, i.e. for $\mbox{\boldmath $\nabla$}_{\xi}\cdot \mbox{\boldmath${J}$} = 0$, all the above quantifiers of information flux vanish, i.e.
\begin{eqnarray}
\frac{D}{Dt}e^{[(1-\beta)R_{\beta(\mathcal{C})}]} =\frac{D~}{Dt}\mathcal{P}_{(\mathcal{C})}=\frac{D~}{Dt}\mathcal{S}_{{\tiny\mbox{vN}}(\mathcal{C})}
=0.
\end{eqnarray}
In addition, given the definitions from Eqs.~\eqref{eqnalexquaz500BB}, \eqref{eqnalexquaz500} and \eqref{eqnzeqnz59}, it is possible to verify that $\mbox{\boldmath $\nabla$}_{\xi}\cdot\mathbf{w}$ is proportional to 
\begin{equation}
\sum_{\eta=1}^{\infty} \left(\frac{i\,\hbar}{2}\right)^{2\eta}\frac{1}{(2\eta+1)!} \, \left[\partial_q^{2\eta+1}V\right]\,W^2\,\partial_p\left(\frac{1}{W}\partial_p \right)^{2\eta}W(q,\,p;\,t),
\label{eqnalexquaz5002}
\end{equation}
and therefore, for parity symmetric potentials, $V(q) =V(-q)$, which lead to periodic anharmonic motions with parity defined wave functions, either in $q$- or $p$- spaces, the above volume integrated contributions (cf. Appendix II) to the information fluxes vanish. 

\subsubsection{Vorticity} To complete the above interpretation of the information flow in the Weyl-Wigner phase-space, also derived from elementary differential operations, the two-dimensional vorticity of a generic phase-space velocity, $\mathbf{w}$, is given by
\begin{equation}\label{rotat}
(\mbox{\boldmath $\nabla$}_{\xi}\times {\mathbf{w}})\cdot\hat{\mathbf{z}} = {\partial_q {\mbox{w}}_p} - {\partial_p {\mbox{w}}_q},
\end{equation}
with $\hat{\mathbf{z}}\cdot \mathbf{v}_{\xi(\mathcal{C})}=\hat{\mathbf{z}}\cdot\mathbf{n}=0$.
The vorticity leads to the circulation number, $\Gamma$, described as 
\begin{eqnarray}
\Gamma 
&=&
\frac{1}{2\pi\hbar}\int_{V_{\mathcal{C}}}dV\,\left({\partial_q {\mbox{w}}_p} - {\partial_p {\mbox{w}}_q}\right)
= \frac{1}{2\pi}\oint_{\mathcal{C}}d{\hat{\ell}} \,({\mbox{w}}_p,\, -{\mbox{w}}_q) \cdot\mathbf{n}
=
\frac{1}{2\pi}\oint_{\mathcal{C}}d\varphi \,\hat{\mbox{\boldmath $
\ell$}}\cdot{\mathbf{w}},\label{eqnquaz692DDMM}
\end{eqnarray}
which accounts for the net effect of quantum affected local spinning motions of the continuum phase-space Wigner velocity distribution, where $\hat{\mbox{\boldmath $\ell$}} = \mathbf{v}_{\xi(\mathcal{C})}/|\mathbf{v}_{\xi(\mathcal{C})}|$ (with $\hat{\mbox{\boldmath $\ell$}}\cdot\mathbf{n}=0$), $d\mathtt{V} = dp\,dq/\hbar$ is a dimensionless phase-space volume unity, and $d\hat{\ell} \sim d \varphi$. For unitary vectors, $\vert\mathbf{w}\vert$, one has either $\Gamma = \pm 1$ or $\Gamma=0$.
As supposed, the result remontes the Stokes theorem, through which the flow's circulation is derived from the path integral of the velocity along a closed path.

In the classical limit, where $\mathbf{w} \to \mathbf{v}_{\xi(\mathcal{C})}$, Eq.\eqref{rotat} for $\hat{\mathbf{w}}$ replaced by ${\mathbf{w}}$ can be identified by the negative value of the phase-space coordinate Laplacian of the Hamiltonian, $H$,
\begin{equation}
\left({\partial_q v_{p(\mathcal{C})}} - {\partial_p v_{q(\mathcal{C})}}\right) = -{\partial^2_q H}-{\partial^2_p H} = -\nabla^2_{\xi} H.
\end{equation}
In this case, one could notice that the vorticity can work as an alternative measure of the {\em distance} from the classical harmonic Hamiltonian system, for which the local Laplacian function, $-\nabla^2 \mathcal{H}_{HO}$, is a constant, i.e.
\begin{equation}
\Sigma = \ln\left[\bigg{\vert}\frac{1}{\nabla^2_{\xi} H}\left({\partial_q \mbox{w}_p} - {\partial_p \mbox{w}_q}\right)\bigg{\vert}\right] =\ln\left[\bigg{\vert}\frac{(\mbox{\boldmath $\nabla$}_{\xi}\times \mathbf{w})\cdot\hat{\mathbf{z}}}{\nabla^2_{\xi} H}\bigg{\vert}\right],
\label{eqnalexquaz5002MM}
\end{equation}
which works as quantifier of anharmonicity and non-linearity introduced by quantum fluctuations over the classical background in the form of vortex effects on the phase-space dynamics, which, of course, returns $\Sigma = 0$ for the classical limit.

Furthermore, the vorticity is strictly related to the crossed derivatives, ${\partial \mbox{w}_p}/{\partial q}$ and ${\partial \mbox{w}_q}/{\partial p}$. For this reason, as it shall be derived in the following section, contributions from ${\partial \mbox{w}_p}/{\partial q}$ and ${\partial \mbox{w}_q}/{\partial p}$ can be strictly connected with the description of the hyperbolic stability properties of non-linear Hamiltonian dynamics.

\section{Hyperbolic stability for non-linear dynamics and the phase-space correspondence}

\hspace{1 em} The equilibrium points of a two-dimensional dynamical system correspond to the solutions of an autonomous system of ordinary differential equations which do not change with time. For the classical phase-space dynamics, equilibrium is geometrically defined and mapped by $\dot{\mbox{\boldmath{$\xi$}}} = 0$ (i.e. $v_{q(\mathcal{C})}=v_{p(\mathcal{C})}=0$). It can be straightforwardly extended to an effective quantum description expressed over the quantum related velocity, $\mathbf{w}$, by $\mbox{w}_q=\mbox{w}_p=0$, with the solutions, in both classical and quantum cases, identified by steady states\footnote{These correspond to 
the so-called fixed points when the continuous system is mapped by a correspondent discretized iterative set of equations.}.
The elementary theoretical grounds for the equilibrium stability are based on the examination of the Jacobian matrix, $j(q,\,p)$, which, in the case of a two-dimensional map,
\begin{eqnarray}
\mbox{w}_q &=& J_q(q,\,p;\,t)/W(q,\,p,\,;t) \equiv \mbox{w}_q(q,\,p),\nonumber\\
\mbox{w}_p &=& J_q(q,\,p;\,t)/W(q,\,p,\,;t) \equiv \mbox{w}_p(q,\,p),
\end{eqnarray}
is identified by
\begin{equation}
j (q,\,p) = \left[\begin{array}{rr}
\partial_q \mbox{w}_q(q,\,p) & \partial_p \mbox{w}_q(q,\,p)\\
\partial_q \mbox{w}_p(q,\,p) & \partial_p \mbox{w}_p(q,\,p)
\end{array}\right].
\end{equation}

The Jacobian matrix defines the equilibrium linear stability properties through its eigenvalues when all derivatives are evaluated at the equilibrium point, ${\mbox{\boldmath{$\xi$}}_e}= (q_e,\,p_e)$, obtained from $\mbox{w}_q(q_e,\,p_e)=\mbox{w}_p(q_e,\,p_e)=0$. Asymptotically stable systems have $j(q_e,\,p_e)$ with all the eigenvalues with negative real parts. 
If at least one eigenvalue of $j(q_e,\,p_e)$ has a positive real part, the system is unstable. 
Through a more robust definition, the Jacobian matrix defines the conditions for the so-called hyperbolic equilibrium if all their corresponding eigenvalues have non-zero real parts. 
The hyperbolic equilibrium admits small linear perturbations over the dynamical system equations which, by the way, do not change qualitatively the phase-space portrait corresponding to the steady state configuration \cite{Book,Book2}.
In this case, the local phase-space portrait of a nonlinear system can be mapped by its linearized perturbative version which equivalently accounts for eventual short displacements of the fixed points (cf. the Hartman-Grobman theorem \cite{HG,HG2}).
Conversely, several types of non-hyperbolic equilibrium patterns result into local bifurcations which may change stability, suppress the fixed point features, or even split them into several equilibrium points.
If at least one eigenvalue of the Jacobian matrix at equilibrium points, $j(q_e,\,p_e)$, has a zero real part, then the equilibrium is not hyperbolic.

Given the above properties, for two-dimensional systems, the hyperbolic equilibrium and stability conditions can be naturally stratified into subclassifications from trace, ${Tr}[\dots]$, and determinant, ${Det}[\dots]$, values of the Jacobian matrix.
Topologically, through the observation of the vector field distribution, $\mathbf{w}(q,\,p)$, at ${\mbox{\boldmath{$\xi$}}_e}= (q_e,\,p_e)$, one can identify: $i)$ saddle points, which correspond to unstable configurations, for real eigenvalues with opposite signs; $ii)$ divergent (unstable) nodes, for both real eingenvalues with positive signs; $iii)$ convergent (stable) nodes, for both real eingenvalues with negative signs; divergent (unstable) focus, for both complex eingenvalues with positive real parts; and $iv)$ convergent (stable) focus, for both complex eingenvalues with negative real parts.

Firstly, one notices that focus and node stabilities are defined by $j(q_e,\,p_e)$ properties such that
\begin{eqnarray}\label{ssss2}
Det[j(q_e,\,p_e)] &=& \partial_q \mbox{w}_q\big{\vert}_{{\mbox{\tiny\boldmath{$\xi$}}_e}}\,\partial_p \mbox{w}_p\big{\vert}_{{\mbox{\tiny\boldmath{$\xi$}}_e}}- \partial_p \mbox{w}_q\big{\vert}_{{\mbox{\tiny\boldmath{$\xi$}}_e}}\,\partial_q \mbox{w}_p\big{\vert}_{{\mbox{\tiny\boldmath{$\xi$}}_e}} > 0,
\end{eqnarray}
with
\begin{eqnarray}
Tr[j(q_e,\,p_e)] &=& \mbox{\boldmath $\nabla$}_{\xi}\cdot\mathbf{w}\big{\vert}_{{\mbox{\tiny\boldmath{$\xi$}}_e}} > 0 \qquad \to \mbox{instability},\nonumber\\
Tr[j(q_e,\,p_e)] &=& \mbox{\boldmath $\nabla$}_{\xi}\cdot\mathbf{w}\big{\vert}_{{\mbox{\tiny\boldmath{$\xi$}}_e}} < 0 \qquad \to \mbox{stability}.
\end{eqnarray}
Likewise, saddle points occur for
\begin{eqnarray}\label{ssss4}
Det[j(q_e,\,p_e)] &=& \partial_q \mbox{w}_q\big{\vert}_{{\mbox{\tiny\boldmath{$\xi$}}_e}}\,\partial_p \mbox{w}_p\big{\vert}_{{\mbox{\tiny\boldmath{$\xi$}}_e}}- \partial_p \mbox{w}_q\big{\vert}_{{\mbox{\tiny\boldmath{$\xi$}}_e}}\,\partial_q \mbox{w}_p\big{\vert}_{{\mbox{\tiny\boldmath{$\xi$}}_e}} < 0.
\end{eqnarray}

Similarly, focus and nodes are separated by $\Delta[j] = Tr[j]^2 - 4 Det[j] = 0$ as
\begin{eqnarray}
\Delta[j(q_e,\,p_e)] && > 0 \qquad \mbox{for nodes},\nonumber\\
\Delta[j(q_e,\,p_e)] && < 0 \qquad \mbox{for focus}.
\end{eqnarray}

From the above analysis and from the Wigner flux tools obtained in the previous section, since $\mbox{\boldmath $\nabla$}_{\xi}\cdot\mathbf{w} =0$ for the classical limit, one can notice that the presence of focus and node sources in the Wigner velocity field is strictly related to the quantum pattern of the Wigner flow.
The Wigner flow vorticity, ${(\mbox{\boldmath $\nabla$}_{\xi}\times \mathbf{w})\cdot\hat{\mathbf{z}}}={\partial_q {\mbox{w}}_p} - {\partial_p {\mbox{w}}_q}$, emerges from the off-diagonal contributions of the Jacobian matrix, $j(q,\,p)$.

Vorticity indeed complements the above content encompassed by the hyperbolic stability analysis. One can notice for instance that, by intrinsically rotating the velocity vector field components by $\Theta = \pi/2\,\, rads$ (clockwise direction cf. Eqs.~\eqref{ads1}-\eqref{ads2} in the following), one gets $\mathbf{w} = (\mbox{w}_q,\,\mbox{w}_p)$ replaced by $\tilde{\mathbf{w}} = (\mbox{w}_p,\,-\mbox{w}_q)$ such that $Det[j(q_e,\,p_e)] = Det[\tilde{j}(q_e,\,p_e)]$, with $\tilde{j}(q_e,\,p_e)$ computed in terms of $\tilde{\mathbf{w}}$. Likewise, one has $\mbox{\boldmath $\nabla$}_{\xi}\cdot\mathbf{w} = (\mbox{\boldmath $\nabla$}_{\xi}\times\tilde{\mathbf{w}})\cdot\hat{\mathbf{z}}$ and $\mbox{\boldmath $\nabla$}_{\xi}\cdot\tilde{\mathbf{w}} = (\mbox{\boldmath $\nabla$}_{\xi}\times{\mathbf{w}})\cdot\hat{\mathbf{z}}$. 

Of course, the jacobian determinant is invariant under generic (intrinsic) $\Theta$-rotations of the velocity vector field components,
\begin{eqnarray}
\label{ads1}\mbox{w}^{\Theta}_q (q,\,p) &=& \mbox{w}_q (q,\,p)\, \cos(\Theta) + \mbox{w}_p (q,\,p)\,\sin(\Theta),\\
\label{ads2}\mbox{w}^{\Theta}_p (q,\,p) &=& \mbox{w}_p (q,\,p)\, \cos(\Theta) - \mbox{w}_p (q,\,p)\,\sin(\Theta),
\end{eqnarray}
such that $Det[j(q_e,\,p_e)] = Det[\tilde{j}^{\Theta}(q_e,\,p_e)]$ and therefore the boundaries for the identification of focus/node and saddle points are not affected by that. More interestingly, from Eqs.~\eqref{ads1}-\eqref{ads2}, one has an additional invariant quantity under $\Theta-rotations$, 
\begin{equation}
\left(\mbox{\boldmath $\nabla$}_{\xi}\cdot\mathbf{w}^{\Theta}\right)^2 + \left[(\mbox{\boldmath $\nabla$}_{\xi}\times \mathbf{w}^{\Theta})\cdot\hat{\mathbf{z}}\right]^2=\left(\mbox{\boldmath $\nabla$}_{\xi}\cdot\mathbf{w}\right)^2 + \left[(\mbox{\boldmath $\nabla$}_{\xi}\times \mathbf{w})\cdot\hat{\mathbf{z}}\right]^2,
\end{equation}
which is equivalent to
\begin{equation}
Tr[j^{\Theta}]^2 + \left[(\mbox{\boldmath $\nabla$}_{\xi}\times \mathbf{w}^{\Theta})\cdot\hat{\mathbf{z}}\right]^2=Tr[j]^2 + \left[(\mbox{\boldmath $\nabla$}_{\xi}\times \mathbf{w})\cdot\hat{\mathbf{z}}\right]^2,
\end{equation}
and from determinant invariance,
\begin{equation}
\Delta[j^{\Theta}] + \left[(\mbox{\boldmath $\nabla$}_{\xi}\times \mathbf{w}^{\Theta})\cdot\hat{\mathbf{z}}\right]^2=\Delta[j] + \left[(\mbox{\boldmath $\nabla$}_{\xi}\times \mathbf{w})\cdot\hat{\mathbf{z}}\right]^2,
\end{equation}
which modulates the conversion of nodes into focus and {\em vice-versa} by intrinsically rotating the phase-space velocity vector field components. The $\Theta$-angle can also be read as the driver of the continuous transition between vorticity and quantumness, namely between $\mathbf{w}\cdot \mathbf{n}\neq 0$ and $\mbox{\boldmath $\nabla$}_{\xi}\cdot\mathbf{w}^{\Theta}\neq 0$ which orthogonally complement each other without affecting the hyperbolic equilibrium structure.

Although it has been discussed in the context of the phase-space vector field interpretation, the problem of limit-cycles and rotated vector fields in the plane, and systematic methods for determining the existence and location
of the cycles of a given second-order system of ordinary differential equations and the limiting sets of non-periodic motions has transversed the last century \cite{Duff51}.
 
As it shall be noticed in the following, the combination of the hyperbolic stability quantifiers with the Wigner flow tools are helpful in quantifying the quantum fluctuations and distinguishing them from non-linear effects when they are addressed to non-linear Hamiltonian systems which, as emphasized, are not systematically accessible through of the Schr\"odinger picture .

\subsection{Phase-space non-linear Hamiltonian dynamics}

Assuming an extended version of the WW framework \cite{2021A,2021B,2021C}, the above identified symplectic structure, which is subjacent to the equilibrium properties of non-linear Hamiltonian systems, can be recovered by the phase-space equations of motion, when they are driven by a Hamiltonian constraint, 
\begin{equation}
H^{W}(q,\,p) = K(p) + V(q),
\label{eqnnlh}
\end{equation}
where, evidently, $\partial^2 H^{W} / \partial q \partial p = 0$, and $K(p)$ and $V(q)$ are arbitrary functions of $p$ and $q$, respectively.

In spite of being analytically based on Hilbert spaces and operators -- usually implemented through the Schr\"odinger equation -- the quantum mechanics resolution of a Hamiltonian system through an {\em eigen}system, $H^{W}\, \psi_n = E_n\, \psi_n$ is not operative for generic non-polinomial functions of $p$ and $q$, $K(p)$ and $V(q)$.
Otherwise, the dynamics described by a {\em Hamiltonian constraint}, as opposed to a Hamiltonian function, recovers the classical mechanics structure geometrically defined on symplectic manifolds.
In this case, the quantum mechanics phase-space framework can be described by means of a fluid analogy through which a probability flux continuity equation provides definitive solutions for quantum ensembles, as opposed to quantum states.

For more enhanced Hamiltonians like those ones from Eq.~\eqref{eqnnlh}, the equivalent continuity equation is cast in the form of Eq.~\eqref{eqnalexquaz51}, with the corresponding Wigner currents now given by \cite{2021A,2021B,2021C}
\begin{equation}
J_q(q,\,p;\,t) = +\sum_{\eta=0}^{\infty} \left(\frac{i\,\hbar}{2}\right)^{2\eta}\frac{1}{(2\eta+1)!} \, \left[\partial_p^{2\eta+1} K(p)\right]\,\partial_q^{2\eta}W(q,\,p;\,t),
\label{eqnalexquaz500BB33}
\end{equation}
and
\begin{equation}
J_p(q,\,p;\,t) = -\sum_{\eta=0}^{\infty} \left(\frac{i\,\hbar}{2}\right)^{2\eta}\frac{1}{(2\eta+1)!} \, \left[\partial_q^{2\eta+1} V(q)\right]\,\partial_p^{2\eta}W(q,\,p;\,t),\label{eqnalexquaz500CC}
\end{equation}
which correspond to the phase-space velocity components, $\mbox{w}_q$ and $\mbox{w}_p$, weighted by the statistical quantum ensemble described by $W\equiv W(q,\,p;\,t)$, i.e. $J_{q(p)} =\mbox{w}_{q(p)}\,W$.

In order to account for quantum corrections coupled with non-linear effects, sometimes non-factorable, a dimensionless version of $H^{W}(q,\,p)$ from Eq.~\eqref{eqnnlh} cast in the form of
\begin{equation}\label{eqnnlhadm}
\mathcal{H}(x,\,k) = \mathcal{K}(k) + \mathcal{V}(x),
\end{equation} 
can be more conveniently considered along the manipulation of the information flux quantifiers.
Eq.~\eqref{eqnnlhadm} is written in terms of dimensionless variables, $x = \left(m\,\omega\,\hbar^{-1}\right)^{1/2} q$ and $k = \left(m\,\omega\,\hbar\right)^{-1/2}p$, such that $\mathcal{H} = (\hbar \omega)^{-1} H$, $\mathcal{V}(x) = (\hbar \omega)^{-1} V\left(\left(m\,\omega\,\hbar^{-1}\right)^{-1/2}x\right)$ and $\mathcal{K}(k) = (\hbar \omega)^{-1} K\left(\left(m\,\omega\,\hbar\right)^{1/2}k\right)$.
In this configuration, one should notice that a mass scale parameter, $m$, and an artificial angular frequency, $\omega$,
are helpful in recasting the Wigner function into a dimensionless form, $\mathcal{W}(x, \, k;\,\omega t) \equiv \hbar\, W(q,\,p;\,t)$, where $dp\,dq$ integrations are replaced by $\hbar\, dx\,dk$ ones, with Wigner currents now written in the form of $\mathcal{J}_x(x, \, k;\,\omega t)$ and $\mathcal{J}_k(x, \, k;\,\omega t)$, so as to give, $\omega\, \partial_x\mathcal{J}_x \equiv \hbar\, \partial_q J_q(q,\,p;\,t)$ and $\omega \,\partial_k\mathcal{J}_k\equiv \hbar \,\partial_p J_p(q,\,p;\,t)$, which finally can all be recast in the form of \cite{NossoPaper,NossoPaper19}
\begin{eqnarray}\label{eqnalexDimW}
\mathcal{W}(x, \, k;\,\tau) &=& \pi^{-1} \int^{+\infty}_{-\infty} \hspace{-.35cm}dy\,\exp{\left(2\, i \, k \,y\right)}\,\psi(x - y;\,\tau)\,\psi^{\ast}(x + y;\,\tau),
\end{eqnarray}
with $y = \left(m\,\omega\,\hbar^{-1}\right)^{1/2} s$ and $\tau = \omega t$,
\begin{eqnarray}\label{eqnalexDimW}
\label{eqnimWA}\mathcal{J}_x(x, \, k;\,\tau) &=& +\sum_{\eta=0}^{\infty} \left(\frac{i}{2}\right)^{2\eta}\frac{1}{(2\eta+1)!} \, \left[\partial_k^{2\eta+1}\mathcal{K}(k)\right]\,\partial_x^{2\eta}\mathcal{W}(x, \, k;\,\tau),\\
\label{eqnimWB}\mathcal{J}_k(x, \, k;\,\tau) &=& -\sum_{\eta=0}^{\infty} \left(\frac{i}{2}\right)^{2\eta}\frac{1}{(2\eta+1)!} \, \left[\partial_x^{2\eta+1}\mathcal{V}(x)\right]\,\partial_k^{2\eta}\mathcal{W}(x, \, k;\,\tau),
\end{eqnarray}
from which, according to Eq.~\eqref{eqnalexquaz51}, one finally has
\begin{equation}
{\partial_{\tau} \mathcal{W}} + {\partial_x \mathcal{J}_x}+{\partial_k \mathcal{J}_k} = {\partial_{\tau} \mathcal{W}} + \mbox{\boldmath $\nabla$}_{\xi}\cdot\mbox{\boldmath $\mathcal{J}$} =0,
\end{equation}
for the associated phase-space coordinates now identified by $\mbox{\boldmath $\xi$} = (x,\,k)$, and with a dimensionless time correspondently defined by $\tau = \omega t$.

\subsubsection{Stationarity} From Eq.~\eqref{eqnalexquaz51}, the explicit form of the stationarity quantifier is straightforwardly given by \cite{2021A}
\begin{equation} \label{helps}
\partial_{\tau} \mathcal{W} = \sum_{\eta=0}^{\infty}\frac{(-1)^{\eta}}{2^{2\eta}(2\eta+1)!} \, \left\{
\left[\partial_x^{2\eta+1}\mathcal{V}(x)\right]\,\partial_k^{2\eta+1}\mathcal{W}
-
\left[\partial_k^{2\eta+1}\mathcal{K}(k)\right]\,\partial_x^{2\eta+1}\mathcal{W}
\right\},\end{equation}
which implicitly accounts for all the contributions from quantum fluctuations ($\mathcal{O}(\hbar^{2\eta})$).

\subsubsection{Classicality} By following a similar structure, with the dimensionless phase-space velocity identified by $\mbox{\boldmath$v$}_{\xi(\mathcal{C})} = \dot{\mbox{\boldmath $\xi$}} = (\dot{x},\,\dot{k})\equiv ({\partial_k \mathcal{H}},\,-{\partial_x \mathcal{H}})$, and with a dimensionless quantum associated velocity parameterized by $\mbox{\boldmath$w$}$, with $\mbox{\boldmath $\mathcal{J}$} = \mbox{\boldmath$w$}\,\mathcal{W}$,
any local discrepancy from the classical regime is identified by $\mbox{\boldmath $\nabla$}_{\xi} \cdot \mbox{\boldmath$w$}\neq 0$, with the Liouvillianity quantifier given by 
\begin{equation}\label{csas}
\mbox{\boldmath $\nabla$}_{\xi} \cdot \mbox{\boldmath$w$} = \sum_{\eta=1}^{\infty}\frac{(-1)^{\eta}}{2^{2\eta}(2\eta+1)!}
\left\{
\left[\partial_k^{2\eta+1}\mathcal{K}(k)\right]\,
\partial_x\left[\frac{1}{\mathcal{W}}\partial_x^{2\eta}\mathcal{W}\right]
-
\left[\partial_x^{2\eta+1}\mathcal{V}(x)\right]\,
\partial_k\left[\frac{1}{\mathcal{W}}\partial_k^{2\eta}\mathcal{W}\right]
\right\}, ~~~\end{equation}
where the $\eta$ contributions are running from $\eta = 1$, since for $\eta =0$ one has the classical limit given by $\mbox{\boldmath $\nabla$}_{\xi} \cdot \mbox{\boldmath$w$} =0$. 

Hence, the quantifiers for stationary and Liouvillian regimes, $\mbox{\boldmath $\nabla$}_{\xi}\cdot\mbox{\boldmath $\mathcal{J}$}$ and $\mbox{\boldmath $\nabla$}_{\xi} \cdot \mbox{\boldmath$w$}$, provide the essential information content for distinguishing quantum from classical regimes as well as their corresponding stationarity.

\subsubsection{Vorticity} Likewise, by following the dimensionless correspondence of Eqs.~\eqref{rotat} and \eqref{eqnalexquaz5002MM}, one has
\begin{eqnarray}
\Sigma &=& \ln\left[\bigg{\vert}\frac{1}{\partial_x^{2}\mathcal{V}(x)+\partial_k^{2}\mathcal{K}(k)} \times\right.\\
&&\left.\sum_{\eta=1}^{\infty}\frac{(-1)^{\eta}}{2^{2\eta}(2\eta+1)!}
\left\{
\partial_k\left\{\left[\partial_k^{2\eta+1}\mathcal{K}(k)\right]\,
\frac{1}{\mathcal{W}}\partial_x^{2\eta}\mathcal{W}\right\}
+
\partial_x\left\{\left[\partial_x^{2\eta+1}\mathcal{V}(x)\right]\,
\frac{1}{\mathcal{W}}\partial_k^{2\eta}\mathcal{W}\right\}
\right\}\bigg{\vert}\right].\nonumber
\label{eqnalexquaz5002NN}
\end{eqnarray}

Contributions to the vorticity pattern also vanish in the classical limit (i.e. for $\eta = 0$), therefore, the classical phase-space Wigner flow is irrotational and divergenceless.

\subsubsection{Hyperbolic Stability} According to the above preliminar discussion, besides the trace of the Jacobian matrix obtained which is equivalent to the Liouvillianity/classicality quantifier from Eq.~\eqref{csas} given by $\mbox{\boldmath $\nabla$}_{\xi} \cdot \mbox{\boldmath$w$}$, the Jacobian determinant according to Eqs.~\eqref{ssss2} and \eqref{ssss4} has a dimensionless form given by
\begin{eqnarray}\label{ssss2m}
Det[j(x,\,k)] &=& \partial_x w_x\,\partial_k w_k- \partial_k w_x\,\partial_x w_k ,
\end{eqnarray}
with
\begin{eqnarray}
\partial_x w_x &=&+\sum_{\eta=1}^{\infty}\frac{(-1)^{\eta}}{2^{2\eta}(2\eta+1)!}
\left\{
\left[\partial_k^{2\eta+1}\mathcal{K}(k)\right]\,
\partial_x\left[\frac{1}{\mathcal{W}}\partial_x^{2\eta}\mathcal{W}\right]
\right\},
\end{eqnarray}
\begin{eqnarray}
\partial_k w_k &=&-\sum_{\eta=1}^{\infty}\frac{(-1)^{\eta}}{2^{2\eta}(2\eta+1)!}
\left\{
\left[\partial_x^{2\eta+1}\mathcal{V}(x)\right]\,
\partial_k\left[\frac{1}{\mathcal{W}}\partial_k^{2\eta}\mathcal{W}\right]
\right\},
\end{eqnarray}
\begin{eqnarray}
\partial_k w_x &=&+\sum_{\eta=1}^{\infty}\frac{(-1)^{\eta}}{2^{2\eta}(2\eta+1)!}
\left\{
\partial_k\left\{\left[\partial_k^{2\eta+1}\mathcal{K}(k)\right]\,
\frac{1}{\mathcal{W}}\partial_x^{2\eta}\mathcal{W}\right\}
\right\},
\end{eqnarray}
and\begin{eqnarray}
\partial_x w_k &=&-\sum_{\eta=1}^{\infty}\frac{(-1)^{\eta}}{2^{2\eta}(2\eta+1)!}
\left\{
\partial_x\left\{\left[\partial_x^{2\eta+1}\mathcal{V}(x)\right]\,
\frac{1}{\mathcal{W}}\partial_k^{2\eta}\mathcal{W}\right\}
\right\},
\end{eqnarray}
which must be evaluated at the equilibrium points corresponding to $w_x(x_e,\,k_e)=w_k(x_e,\,k_e)=0$ as to provide the pattern of stability according to the hyperbolic conditions.

\subsection{Harper-like systems}

\begin{figure}[h]
\includegraphics[scale=0.5]{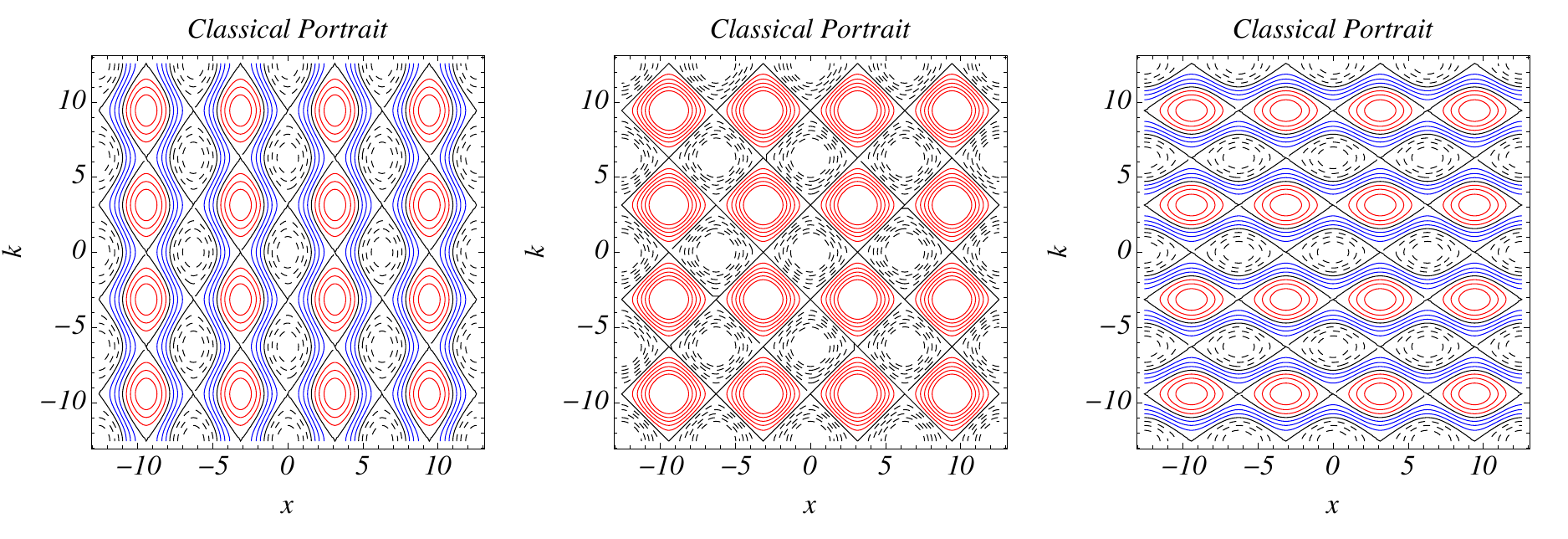}
\renewcommand{\baselinestretch}{.85}
\caption{\footnotesize
(Color online) Classical portrait of Harper-like lattice. Phase-space trajectories and corresponding lattice designs are for $ \mbox{Max}\{\nu^2-1,0\} < \vert\epsilon\vert < \nu^2+1$ corresponding to closed trajectories for $\epsilon > 0$ (black dashed lines), and for $\epsilon < 0$ (red thin lines), and with $0 < \vert\epsilon\vert < \nu^2-1$ corresponding to opened trajectories (blue thick lines), when they exist. The limiar (opened-closed) value is given by $\vert\epsilon\vert = \nu^2 -1$. The plots are for $\nu^2=2$ (first plot), with $\vert\epsilon\vert = 5/2,\, 2,\,3/2,\,\dots,\,0$, $\nu^2=1$ (second plot), with $\vert\epsilon\vert = 5/4,\, 1,\,3/4,\,\dots,\,0$ and $\nu^2=1/2$ (third plot), with $\vert\epsilon\vert = 5/4,\, 1,\,3/4,\,\dots,\,0$. Plots are similar to those from Ref. \cite{2021A}, with the classical energy identified by $\mathcal{H}_H \to \epsilon$, and with arbitrary values for phenomenological parameter, $\nu$.}
\label{eqnHarperHarper}
\end{figure}

Finally, in order to fit the non-linear properties discussed above, one can consider, for instance, the Hamiltonian described by
\begin{equation}\label{eqnHamHarper01dim}
\mathcal{H}_H(x,\,k)= \cos(k) +\nu^2 \cos(x),
\end{equation}
which, besides exhibiting periodic analytical solutions corresponding to phase-space closed orbits as depicted in Fig. \ref{eqnHarperHarper}, corresponds to a resumed classical version of Harper-like Hamiltonians \cite{Harper01,Harper01BB,Harper01CC,Harper02,PRL-Harper} with an anisotropic behavior parameterized by $\nu^2$.
Harper-like models \cite{NatHarper,RMPRMP,PRA-Harper19} were introduced through a $1$-dim Hamiltonian parameterization of nearest-neighbor couplings modulated by a senoidal function modulation, with the on-site energies resumed by the Hamiltonian operator working on the quantum state, $\psi_n$, as
\begin{equation}\label{eqnHamHarper00}
\mathcal{H} \psi_n = -A_k (e^{+i\vartheta}\,\psi_{n+1} + e^{-i\vartheta}\,\psi_{n-1}) - A_x\,\cos(2\pi
\alpha\, n +\theta) \psi_n,\end{equation}
where $A_{x}$ and $A_{k}$ are the parameters for coupling magnitude and modulation \cite{PRL-Harper}, respectively, and phases $\theta$ and $\vartheta$ are associated to the wave number in $2$-dim. 

To recover the classical form from Eq.~\eqref{eqnHamHarper01dim}, one notices that quantum number displacements featured by $\psi_{n\pm1}$ contributions to (\ref{eqnHamHarper00}) corresponds to localized states in the nearest-neighbor sites.
Therefore, since one can also identifies that $\psi_{n\pm1}\sim \psi({x\pm a}) \equiv \exp[\pm i\,k\,a]\psi(x)$, which is parameterized by translation operations, i.e. $\exp[\pm i\,k\,a]$, the momentum operator, $\pm k$ (for a coordinate correspondence given by $(x,\,a)\to(2\pi\alpha n, 2\pi\alpha)$), once applied into $\psi_{n}\equiv\psi(x)$, so as to return $\psi_{n\pm1}\psi({x\pm a})$, straightforwardly admits a semi-classical representation in the form of Eq.~\eqref{eqnHamHarper01dim}.
For Eq.~(\ref{eqnHamHarper01dim}) obtained from (\ref{eqnHamHarper00}) \cite{Harper02} by setting $A_{x}/A_{k}=\nu^2$, with an arbitrary phenomenological parameter, $\nu$, the position and momentum coordinate operators, $\hat{x}$ and $\hat{k}$, can be read as non-commutative variables satisfying $[\hat{x},\hat{k}] = i\,2\pi\alpha$, where $2\pi\alpha$ plays the role of the dimensionless reduced Planck constant ($\sim 1$).

By examining the classical-to-quantum transition from a statistical perspective, one finds that the classical analog of the second-order moments of position and momentum -- parameters associated with Heisenberg's uncertainty principle -- should consistently obey the same constraints as their quantum counterparts. This assumption is effectively isomorphic to the framework of Gaussian quantum mechanics, in the sense that it yields equivalent statistical predictions. Consequently, replacing the Wigner distributions with Gaussian distributions represents the most natural procedure for identifying the classical-to-quantum transition in Hamiltonian systems that approximately follow a normal distribution in phase space.

Hence, considering Gaussian ensembles (cf. the Appendix III), now driven by $\mathcal{K}(k)=\cos(k)$ and $\mathcal{V}(x)=\nu^2\cos(x)$, after some straightforward mathematical manipulations involving Eqs.~\eqref{eqnt111} and \eqref{eqnt222}, Eqs.~\eqref{eqnimWA4} and \eqref{eqnimWB4}, can be replaced by \cite{2021A}
\begin{eqnarray}
\label{eqnimWA4CC}\partial_x\mathcal{J}_x(x, \, k) &=& +2 \,\sin\left(x\right)\,\sinh\left(\gamma^2\,x\right)\,\exp[-\gamma^2 /4]\,\mathcal{G}_{\gamma}(x, \, k)\,
,\\
\label{eqnimWB4CC}\partial_k\mathcal{J}_k(x, \, k) &=& -2 \nu^2\,\sin\left(k\right)\,\sinh\left(\gamma^2\,k\right)\,\exp[-\gamma^2 /4]\,\mathcal{G}_{\gamma}(x, \, k),
\end{eqnarray}
which results from the convergent series expansion resumed by Eqs.~\eqref{eqnimWA3} and \eqref{eqnimWB3}.
The integrated Wigner currents obtained from Eqs.~\eqref{eqnimWA4} and \eqref{eqnimWB4} are thus written as \cite{2021A}
\begin{eqnarray}
\label{eqnimWA4CCD}\mathcal{J}_x(x, \, k) &=& +\frac{\gamma}{2\sqrt{\pi}} \,\sin\left(x\right)\,\exp\left(-\gamma^2\,k^2\right)\,
\left[Erf\left(\gamma(x-1/2)\right)-Erf\left(\gamma(x+1/2)\right)\right],\\
\label{eqnimWB4CCD}\mathcal{J}_k(x, \, k) &=& -\frac{\gamma}{2\sqrt{\pi}} \nu^2\,\sin\left(k\right)\,\exp\left(\gamma^2\,x^2\right)\,
\left[Erf\left(\gamma(k-1/2)\right)-Erf\left(\gamma(k+1/2)\right)\right],
\end{eqnarray}
where $Erf(\dots)$ are the {\em error functions} and the components of the quantum velocities, $\mathbf{w}$, $w_x$ and $w_k$, can be obtained simply by replacing $-k^2 \leftrightarrow +x^2$ in the exponential function of the respective Eqs.~\eqref{eqnimWA4CCD} and \eqref{eqnimWB4CCD}.

\begin{figure}[h!]
\vspace{-1.2 cm}\includegraphics[scale=0.16]{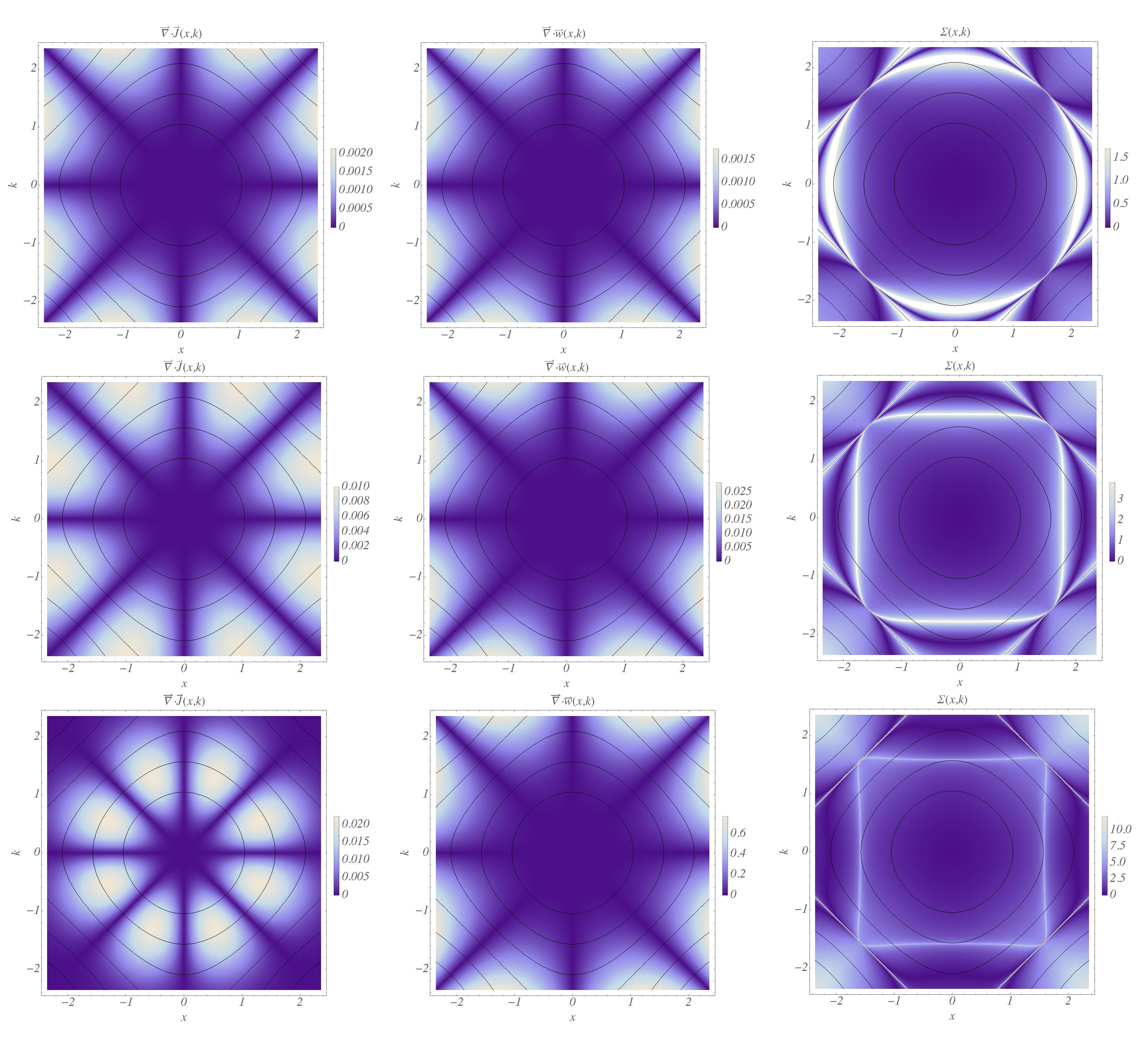}
\renewcommand{\baselinestretch}{.6}
\vspace{-1.2 cm}\caption{\footnotesize
(Color online) Wigner flow quantifier pattern for Gaussian ensembles in the $x - k$ plane.
{\em First column}:
Stationarity quantifier, $\mbox{\boldmath $\nabla$}_{\xi} \cdot \mbox{\boldmath $\mathcal{J}$}$, described according to the background color scheme. The results are for the increasing spreading characteristic of the Gaussian function, from $\gamma =1/4$ (first row), $1/2$ (second row) and $1$ (third row). Peaked Gaussian distributions ($\gamma =1$ in ) localizes the quantum distortions which result into non-stationarity.
{\em Second column}: Liouvillian quantifier, $\mbox{\boldmath $\nabla$}_{\xi} \cdot \mathbf{w}$, depicted through the background color scheme, from darker regions, $\mbox{\boldmath $\nabla$}_{\xi} \cdot \mathbf{w} \sim 0$, to lighter regions, $\mbox{\boldmath $\nabla$}_{\xi} \cdot \mathbf{w} > 0$.
{\em Third column}: Circulation quantifier, $\Sigma$, for Gaussian ensembles which do not exhibit neither vortices nor stagnation points, in a kind of camouflage of the quantum distortions. 
The classical pattern is shown as a collection of black lines.}
\label{eqnHarperHarper04}
\end{figure}

The confront between classical and quantum regimes quantified by stationarity, Liouvillianity, and vorticity quantifiers can be depicted in Fig. \ref{eqnHarperHarper04}, where the associated density plots for $\mbox{\boldmath $\nabla$}_{\xi} \cdot \mbox{\boldmath $\mathcal{J}$}$, $\mbox{\boldmath $\nabla$}_{\xi} \cdot \mathbf{w}$ and $\Sigma$ are identified.	
The corresponding Gaussian Wigner flow pattern plots follow the background color scheme, from which lighter regions correspond to non-vanishing quantum distortion local contributions, and darker regions approximate the classical pattern.
In particular, one can notice how the magnitude of the perturbations evolve with the Gaussian envelop parameter, $\gamma$.
The vorticity quantifier, $\Sigma$, do not exhibit neither vortices nor stagnation points, in a kind of camouflage of the Gaussianized quantum distortions. However, since the classical boundaries do not follow the Gaussian distribution contours, the circulation is highly distorted by quantum contributions far from the equilibrium points at the origin, $(x_e,\,k_e) = (0,\,0)$.

\begin{figure}[h!]
\includegraphics[scale=0.3]{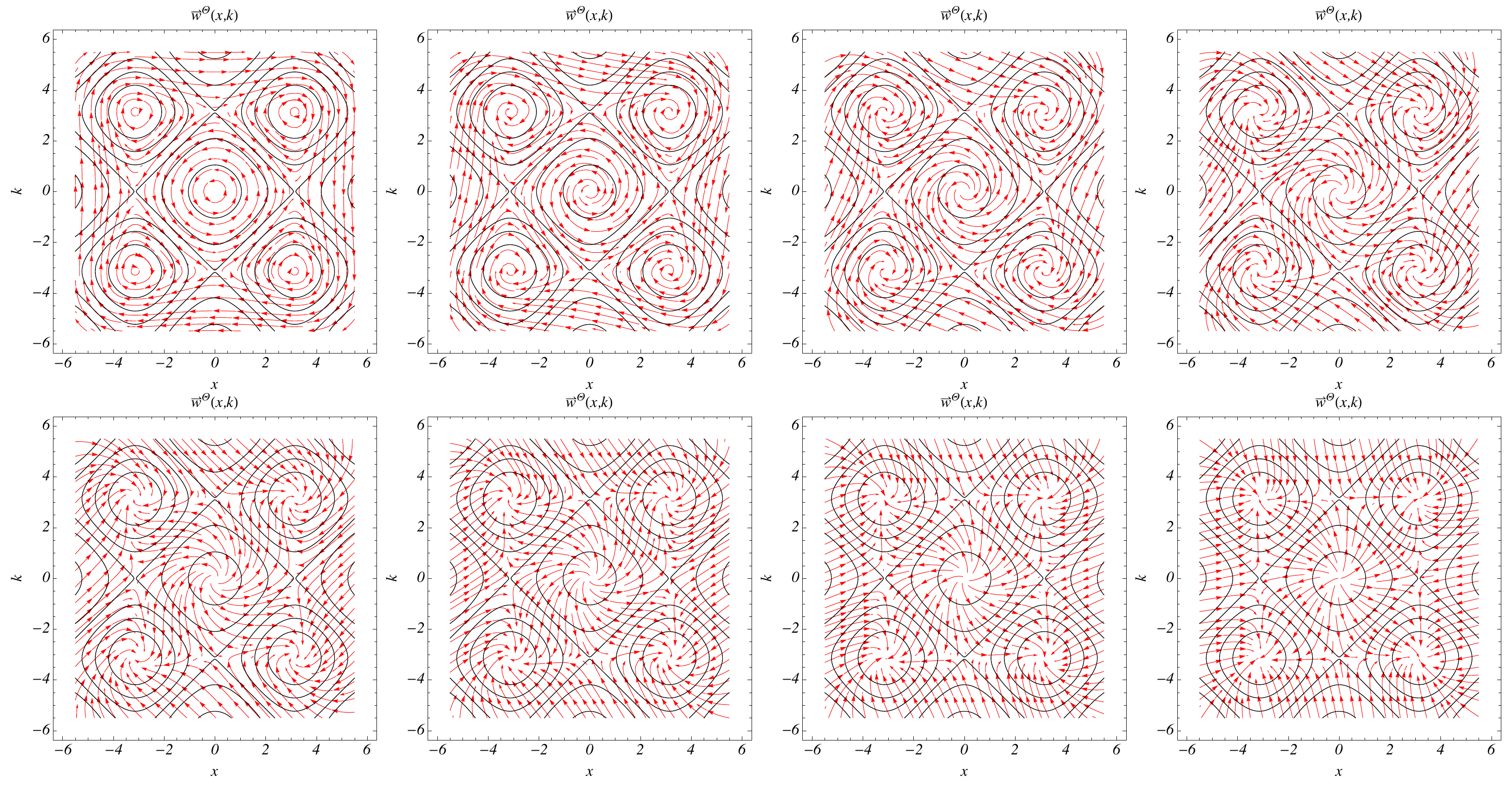}
\renewcommand{\baselinestretch}{.6}
\caption{\footnotesize
(Color online) Velocity vector field pattern, $\mathbf{w}^{\Theta}$, evolving terms of $\Theta$-rotations, as parameterized by the canonical transformations from Eqs.~\eqref{ads1}-\eqref{ads2}, from a divergenceless regime at equilibrium points ($\Theta = 0$) to an almost completely irrotational regime at $\Theta = 7\pi/16$ in the $x - k$ plane (with $\Theta = n \pi/16, with n=0,\,1,\,\dots,\,7$ from left to right, and from top to bottom). 
The classical pattern is shown as a collection of black lines. Results are for $\gamma = 1/4$ -- a choice which qualitatively does not affect the above interpretation}
\label{HHRot}
\end{figure}

The results from Eqs.~\eqref{eqnimWA4CCD} and \eqref{eqnimWB4CCD} also provides the pattern for the hyperbolic equilibrium and stability. The evolution of the velocity vector field pattern, $\mathbf{w}^{\Theta}$, in terms of $\Theta$, as parameterized by the canonical transformations from Eqs.~\eqref{ads1}-\eqref{ads2} is depicted from Fig. \ref{HHRot}, from a divergenceless regime at equilibrium points ($\Theta = 0$) to an almost completely irrotational regime at $\Theta = 7\pi/16$

\begin{figure}[h!]
\includegraphics[scale=0.17]{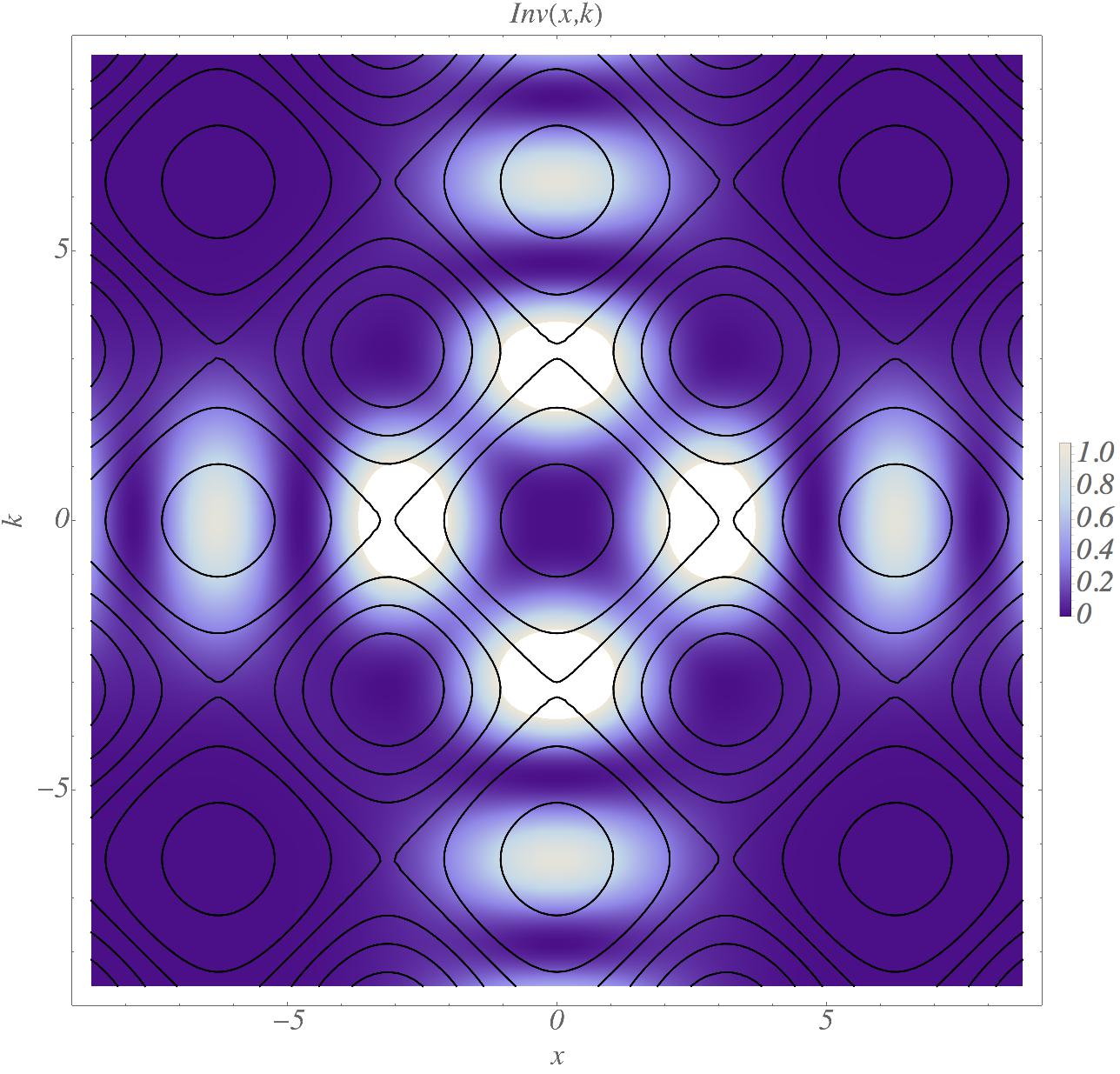}
\includegraphics[scale=0.17]{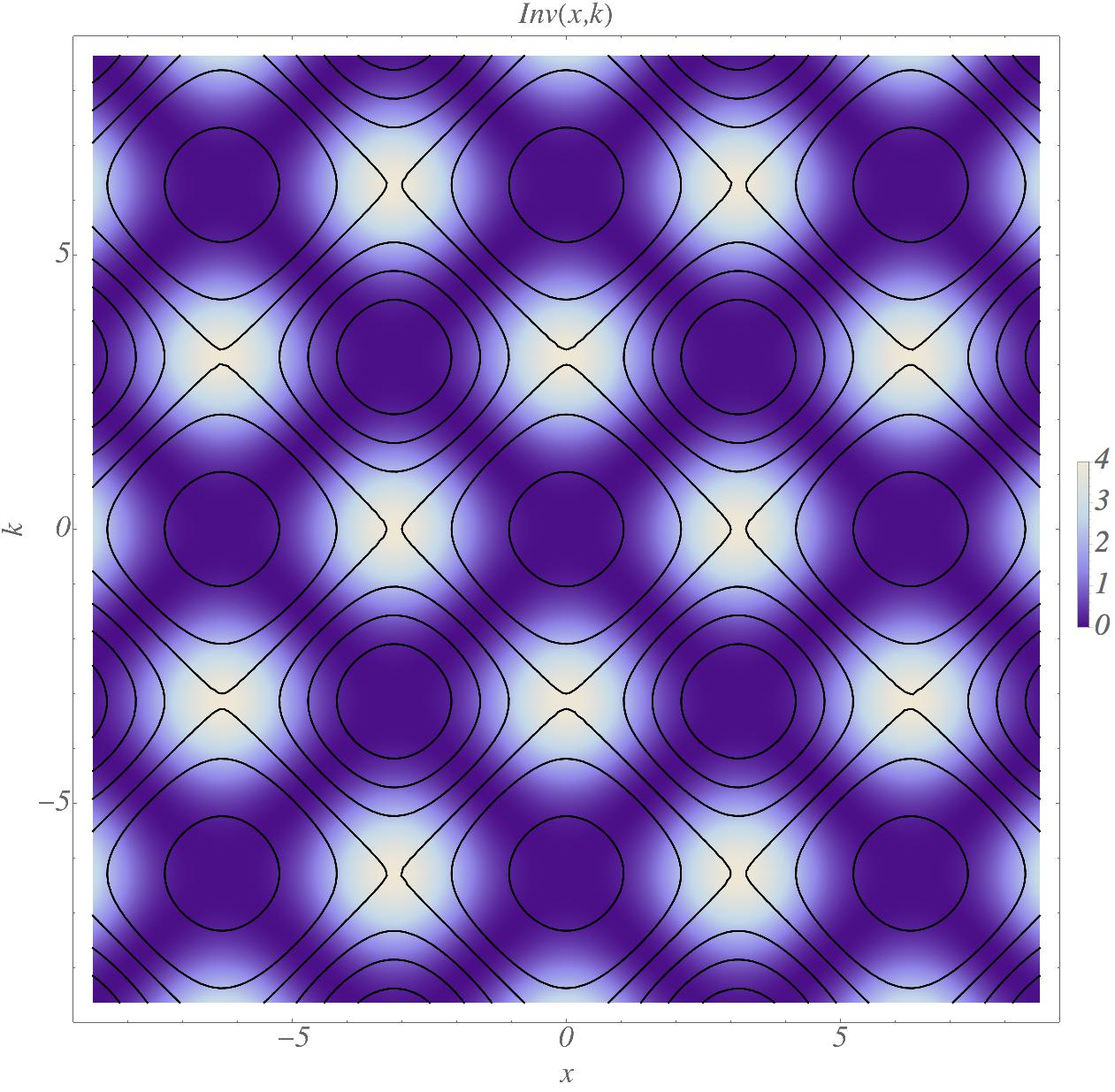}
\includegraphics[scale=0.17]{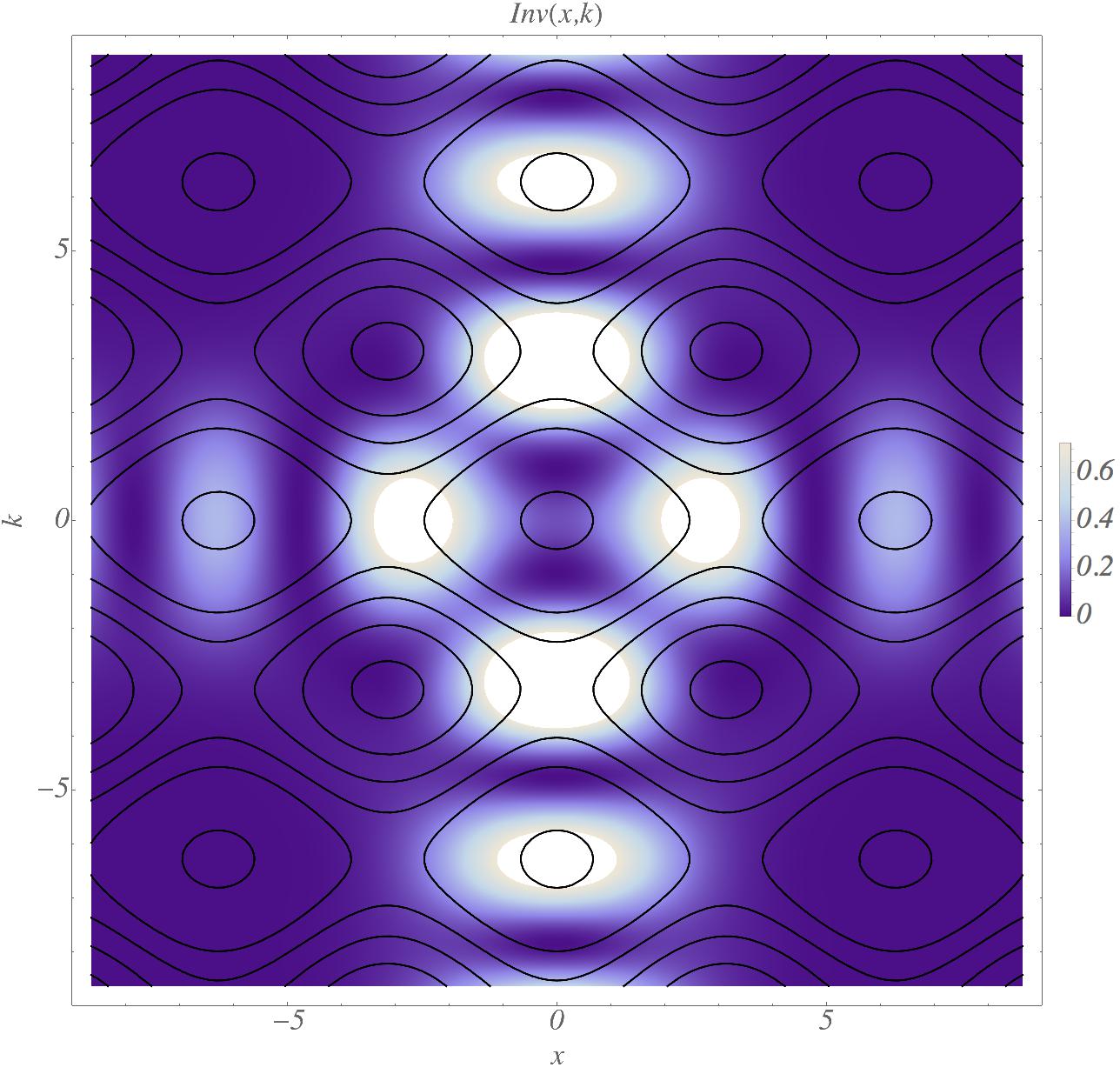}
\includegraphics[scale=0.17]{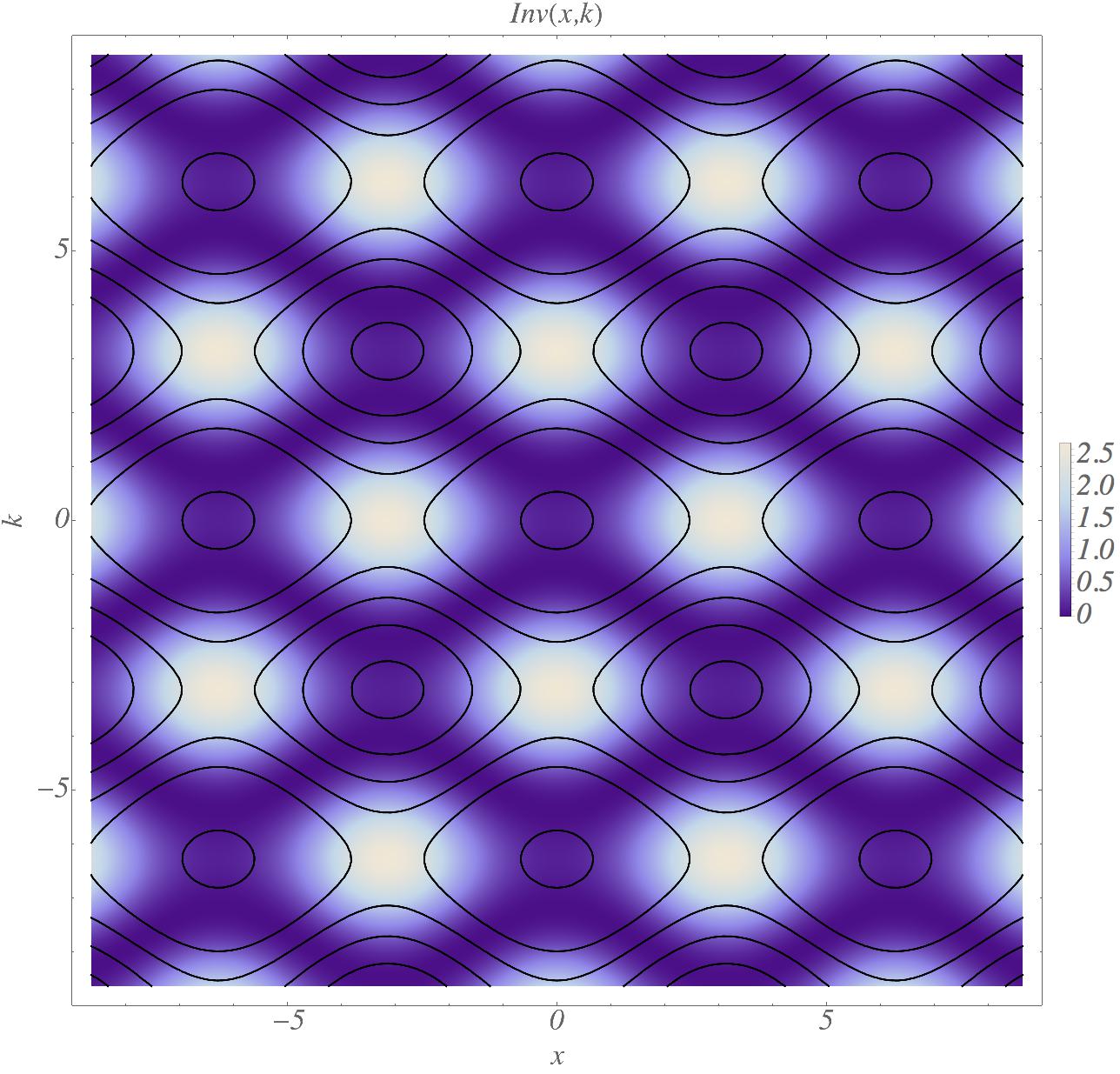}
\renewcommand{\baselinestretch}{.6}
\caption{\footnotesize
(Color online) $\Theta$-rotational invariant quantifier, $Inv(x,\,k)$, for quantum ($\alpha = 1/4$, first column) and classical ($\alpha \to 0$, second column) for isotropic ($\nu=1$, first row) and anisotropic ($\nu=0.8$, second row) regimes. 
The classical pattern is shown as a collection of black lines in both plots.}
\label{Novass}
\end{figure}

However, for the Hamiltonian Eq.~(\ref{eqnHamHarper01dim}), both classical and quantum regimes exhibit stable closed orbit patterns. The complementary behavior between Liouvillian and circulation quantifiers from second and third columns of Fig. \ref{eqnHarperHarper04}, respectively, can be depicted from Fig. \ref{Novass}, in which the $\Theta$-rotation invariant quantifier,
\begin{equation}
Inv(x,\,k) = Tr[j(x,\,k)]^2 +\left[(\mbox{\boldmath $\nabla$}_{\xi}\times \mathbf{w}(x,\,k))\cdot\hat{\mathbf{z}}\right]^2- 4 Det[j(x,\,k)]
\end{equation}
is depicted for the isotropic ($\nu = 1$) configurations of quantum and classical regimes.

In fact, since the equilibrium points are given by $(x_e,\,k_e) = (0,\,0)$, one has closed orbits with $\mbox{\boldmath $\nabla$}_{\xi} \cdot \mathbf{w}(x_e,\,k_e) = Tr[j(x_e,\,k_e)] =0$, and
$$\Delta[j(x_e,\,k_e)] = Tr[j(x_e,\,k_e)]^2 - 4 Det[j(x_e,\,k_e)] = -\frac{4 \pi \nu^2}{\gamma ^2}Erf\left(\frac{\gamma }{2}\right)^2 < 0,$$
$$(\mbox{\boldmath $\nabla$}_{\xi}\times \mathbf{w}(x_e,\,k_e))\cdot\hat{\mathbf{z}}= \frac{(1+\nu^2) \sqrt{\pi}}{\gamma}Erf\left(\frac{\gamma }{2}\right),$$
with the $\Theta$-rotation invariant:
$$Tr[j(x_e,\,k_e)]^2 +\left[(\mbox{\boldmath $\nabla$}_{\xi}\times \mathbf{w}(x_e,\,k_e))\cdot\hat{\mathbf{z}}\right]^2- 4 Det[j(x_e,\,k_e)] = \frac{(1-\nu^2)^2 \pi }{\gamma ^2}Erf\left(\frac{\gamma }{2}\right)^2.$$
Of course, the inclusion of additional analytical anisotropic perturbations, for instance, as those ones observed in the Aubry-Andr\'e-Harper formulation for classifying topological states \cite{PRBPRB2}, should introduce deviations from stable orbit regimes, which probably deserves a deeper analysis in terms of the tools presented in this work.

\section{Conclusion}

In view of the originality and broad scope of the Wigner phase-space formalism for non-linear Hamiltonians of the form $H^{W}(q,\,p) = K(p) + V(q)$, we have developed a consistent geometric mapping of hyperbolic stability conditions into quantum phase-space variables. This was accomplished within a self-contained framework that generalizes classical dynamical systems through the differential geometric structure of Wigner flow metrics. These metrics emerge from a fluid-analogy interpretation of the generalized Wigner current and allow for the systematic extraction of dynamical features such as stationarity, classicality, purity, and vorticity in non-linear quantum systems.

The proposed formalism connects naturally with foundational studies on the geometry of phase-space dynamics. For instance, the work by Liu \cite{Liu2011} discussed the relationship between the continuity equation and the Wigner-Moyal dynamics \cite{Moyal} in thermal equilibrium, establishing the non-uniqueness of the effective phase-space force and proposing alternative flux constructions -- a point that directly supports the present treatment of generalized Wigner currents. Furthermore, recent developments using trajectory-based and machine learning approaches \cite{Liu2021b, Zhou2022} offer efficient computational pathways to implement and explore this formalism in practical simulations.

An additional relevant perspective can be also provided by alternative representations of quantum phase-space dynamics. Although the Weyl-Wigner-Moyal formulation forms the backbone of our construction, it is not the only viable representation. The Husimi formulation \cite{Husimi}, and more generally, quasi-probability distributions such as the Glauber-Sudarshan \cite{Glauber, Sudarshan} and optical tomographic representations \cite{Amosov, Radon, Mancini}, provide smoother, often positive-definite alternatives that may circumvent interpretative issues arising from the negativity of the Wigner function \cite{Ballentine, Carmichael, Callaway}. In this context, the works by Veronez and de Aguiar \cite{Veronez2013, Veronez2016} on the structure and topology of the Husimi flow are noteworthy, showing that flow fields and their stationary points carry nontrivial topological information even in smoothed phase-space formulations. Such approaches complement our results and motivate further exploration of how geometric and topological structures manifest under different quantization schemes.

Within the Wigner framework, specialization to Gaussian ensembles reveals that a broad class of quantum systems can be treated consistently. As an illustrative case, the results for Harper-like models were extended to the Aubry-Andr\'e system via reparameterized linear terms in $H^{W}(q,\,p)$ \cite{Ber2025}. Beyond condensed matter physics, Toda lattice systems \cite{sigma}, which involve Hamiltonians like $K(p) = \cosh(p/p_0)$, also fall naturally within this class. These systems, when viewed through their integrable properties and connections to spectral and string theory \cite{Pasquier}, provide fertile ground for the application of the generalized Wigner flow framework.

The formalism also extends to broader system classes, including population dynamics and stochastic models. Lotka-Volterra-type systems \cite{LV,LV2,2021A,Ber2024} and other stochastic processes \cite{Allen,Grasman} can be cast into effective Hamiltonian form and studied using the similar geometric tools \cite{Veronez2013,Veronez2016,sigma,RPSA-LV,Ber2025}. Here, the inclusion of quantum fluctuations provides new means of analyzing stability, bifurcation, and transition phenomena, extending the applicability of quantum phase-space techniques to fields such as ecology, epidemiology, and evolutionary game theory \cite{PRE-LV,SciRep02,PRE-LV2,Nature01,Nature02}.

Suitably, the framework developed here invites extensions to systems involving both discrete and continuous degrees of freedom, such as spin-boson models, polaritonic states, and nonadiabatic quantum dynamics. Recent efforts \cite{Zhou2022, Liu2025} demonstrate the relevance of composite-variable quantum dynamics in simulating molecular systems and understanding the role of coherence and entanglement. Our formulation, grounded in the geometric structure of quantum flows, can be more properly scrutinized to capture these hybrid dynamics.

In summary, the present work systematically incorporates quantum fluctuations into the stability analysis of non-linear Hamiltonian systems. It establishes a geometrically self-contained framework for identifying topological and dynamical features of phase-space quantum mechanics, and it suggests bridges to complementary approaches in both theoretical and computational domains. The results are promising not only from a foundational viewpoint but also for future applications in complex systems, molecular dynamics, and quantum technologies.

{\em Acknowledgments -- This work is supported by the Brazilian Agency CNPq (Grant No. 301485/2022-4).}

\section*{Appendix I - Properties of the Wigner flow}

Departing from the theorem for the rate of change of the volume integral bounded by a comoving closed surface (cf. Eq.~(10.811) from Ref. \cite{Gradshteyn}) written in terms of a {\em substantial derivative} 
\begin{equation}
\frac{D~}{Dt} \equiv \partial_t + \mathbf{v}_{\xi} \cdot\mbox{\boldmath $\nabla$}_{\xi} 
\label{eqnalexquaz57DDD},
\end{equation}
as \cite{MeuPT,NossoPaper,NossoPaper19} 
\begin{equation}
\frac{D~}{Dt} \int_{V}dV\,{f[W]} \equiv 
\int_{V}dV\,\left(\frac{D}{Dt} + \mbox{\boldmath $\nabla$}_{\xi}\cdot \mathbf{v}_{\xi}\right){f[W]}\label{eqnalexquaz57D},
\end{equation}
the differential form of the conservative properties related to the elementary vector unity of the phase-space, $\mbox{\boldmath $\xi$} \equiv (q,\,p)$, with $dV \equiv dp\,dq$, follows from the analysis of the classical and quantum-like vector velocities, $\mathbf{v}_{\xi}$ and $\mathbf{w}$, and of their associated currents $\mbox{\boldmath${J}$}^{\mathcal{C}}$ and $\mbox{\boldmath${J}$}$.
The corresponding integral forms, which can be derived from modified versions of the continuity equation from (\ref{eqnalexquaz51}) (besides itself) \cite{MeuPT,NossoPaper,NossoPaper19}, emerge from the assumption of a two-dimensional phase-space volume, $V_{_{\mathcal{C}}}$, enclosed by the comoving closed surface, $\mathcal{C}$, which corresponds to the path defined by the classical velocity, $\mathbf{v}_{\xi(\mathcal{C})}$.
In this case, from Eq.~\eqref{eqnalexquaz57D}, with $f[W]$ conveniently replaced by an arbitrary power of the Wigner function, $W^{\beta}$, with the introduction of an auxiliary unitary vector, $\mathbf{n}$, satisfying $\mathbf{n}\cdot\mathbf{v}_{\xi(\mathcal{C})}= 0$, one has
\begin{eqnarray}
 \frac{D~}{Dt}\left(\int_{V_{\mathcal{C}}}dV\, {W}^{\beta}\right)
&=& \int_{V_{\mathcal{C}}}dV\,\left[\frac{D~}{Dt} ({W}^{\beta}) + {W}^{\beta} \mbox{\boldmath $\nabla$}_{\xi}\cdot \mathbf{v}_{\xi(\mathcal{C})}\right]\\
&=&\int_{V_{\mathcal{C}}}dV\,\left[\partial_t ({W}^{\beta}) + \mbox{\boldmath $\nabla$}_{\xi}\cdot(\mathbf{v}_{\xi(\mathcal{C})} {W}^{\beta})\right]\nonumber\\
&=&\int_{V_{\mathcal{C}}}dV\,\left[-\beta{W}^{\beta-1} \mbox{\boldmath $\nabla$}_{\xi}\cdot\mbox{\boldmath${J}$} + \mbox{\boldmath $\nabla$}_{\xi}\cdot(\mathbf{v}_{\xi(\mathcal{C})} {W}^{\beta})\right]\nonumber\\
&=&-\int_{V_{\mathcal{C}}}dV\,\left[(\beta-1){W}^{\beta} \,\mbox{\boldmath $\nabla$}_{\xi}\cdot\mathbf{w} 
+ \mbox{\boldmath $\nabla$}_{\xi}\cdot(\mbox{\boldmath${J}$}{W}^{\beta-1} - \mathbf{v}_{\xi(\mathcal{C})} {W}^{\beta})\right]\nonumber\\
 &=& -\int_{V_{\mathcal{C}}}dV\,(\beta-1){W}^{\beta} \,\mbox{\boldmath $\nabla$}_{\xi}\cdot\mathbf{w} 
- \oint_{\mathcal{C}}d\ell\, {W}^{\beta-1}\left((\mbox{\boldmath${J}$}-\mbox{\boldmath${J}$}^{\mathcal{C}})\cdot \mathbf{n}\right)
\nonumber\\
 &=& -\int_{V_{\mathcal{C}}}dV\,(\beta-1){W}^{\beta} \,\mbox{\boldmath $\nabla$}_{\xi}\cdot\mathbf{w} 
- \oint_{\mathcal{C}}d\ell\, {W}^{\beta-1}\left(\mbox{\boldmath${J}$}\cdot \mathbf{n}\right),\nonumber\label{eqnquaz692}
\end{eqnarray}
with
$$\frac{1}{2\pi}\oint_{\mathcal{C}}\frac{d\ell}{|\mathbf{v}_{\xi(\mathcal{C})}|}\equiv\frac{1}{2\pi}\int_0^{\2\pi}{d\varphi}=1,\qquad(\mbox{closed classical trajectory, $\mathcal{C}$}),$$
and which is converted into an efficient tool for quantifying several aspects related to the flux of information through $\mathcal{C}$ \cite{NossoPaper,NossoPaper19}.

\section*{Appendix II - Information flux continuity equations}

For parity symmetric potentials, $V(q) =V(-q)$, a set of information flux continuity equations obtained in Sec. II can be cast in terms of path integration contributions, in the form of \cite{NossoPaper,NossoPaper19,Meu2018}
\begin{eqnarray}
\frac{D~}{Dt}\sigma_{(\mathcal{C})} + \oint_{\mathcal{C}}d\ell\,\left(\mbox{\boldmath${J}$}\cdot \mathbf{n}\right) &=& 0,\nonumber\\\frac{1}{2\pi\hbar} \frac{D~}{Dt}\mathcal{P}_{(\mathcal{C})} + \oint_{\mathcal{C}}d\ell\, {W}\,\left(\mbox{\boldmath${J}$}\cdot \mathbf{n}\right) &=& 0,\nonumber\\
\frac{D~}{Dt}\mathcal{S}_{{\tiny\mbox{vN}}(\mathcal{C})}-\oint_{\mathcal{C}}d\ell\, \ln|2\pi\hbar{W}|\,\left(\mbox{\boldmath${J}$}\cdot \mathbf{n}\right)&=&0,\nonumber\\
\frac{1}{(2\pi\hbar)^{(\beta-1)}}\frac{D}{Dt}e^{[(1-\beta)R_\beta]}+\oint_{\mathcal{C}}d\ell\, {W}^{\beta-1}\left(\mbox{\boldmath${J}$}\cdot \mathbf{n}\right)&=&0,
\label{eqnquaz692DDEEFF}
\end{eqnarray}
where, for periodic motions with a period $T$, the path integrals along the closed classical path, $\mathcal{C}$, can all be re-parameterized as
\begin{eqnarray}
\oint_{\mathcal{C}}d\ell\, f[W]\left(\mbox{\boldmath${J}$}\cdot \mathbf{n}\right)&=&
\oint_{\mathcal{C}}d\ell\, f[W]\left((\mbox{\boldmath${J}$} - \mbox{\boldmath${J}$}^{\mathcal{C}})\cdot \mathbf{n}\right)\nonumber\\&=&
-\int_0^T dt\,\frac{d\ell}{dt}\, f\left[W(q_{\mathcal{C}}(t),\,p_{\mathcal{C}}(t);\,t)\right]\,|\mathbf{v}_{\xi(\mathcal{C})}|^{-1}\,\left(J_p - J_p^{\mathcal{C}}\right)\,\frac{dq_{\mathcal{C}}}{dt}\nonumber\\&=&
-\int_0^T dt\, f\left[W(q_{\mathcal{C}}(t),\,p_{\mathcal{C}}(t);\,t)\right]\,\Delta J_p(q_{\mathcal{C}}(t),\,p_{\mathcal{C}}(t);\,t)\,\frac{dq_{\mathcal{C}}}{dt},
\label{eqnquaz692DDEEFF}
\end{eqnarray}
where ${d\ell}/{dt} = |\mathbf{v}_{\xi(\mathcal{C})}|$, $J_p - J_p^{\mathcal{C}} \equiv \Delta J_p(q_{\mathcal{C}}(t),\,p_{\mathcal{C}}(t);\,t)$ and $q_{\mathcal{C}}(t)$, and $p_{\mathcal{C}}(t)$ correspond to the classically evolved coordinates.

\section*{Appendix III - Gaussian ensembles}

In order to verify the effectiveness of the above results, Gaussian ensembles can be considered as the input Wigner function written as
\begin{equation}
\mathcal{G}_\gamma(x,\,k) = \hbar \,G_\gamma(q,\,p) = \frac{\gamma^2}{\pi}\, \exp\left[-\gamma^2\left(x^2+ k^2\right)\right],
\end{equation}
where the dependence on the time parameter, $\tau$, is implicitly given in terms of $x\equiv x(\tau)$ and $k\equiv k(\tau)$.

The associated Wigner flow contributions given by
\begin{eqnarray}
\label{eqnimWA2}\partial_x\mathcal{J}_x(x, \, k) &=& +\sum_{\eta=0}^{\infty} \left(\frac{i}{2}\right)^{2\eta}\frac{1}{(2\eta+1)!} \, \left[\partial_k^{2\eta+1}\mathcal{K}(k)\right]\,\partial_x^{2\eta+1}\mathcal{G}_{\gamma}(x, \, k),
\\
\label{eqnimWB2}\partial_k\mathcal{J}_k(x, \, k) &=& -\sum_{\eta=0}^{\infty} \left(\frac{i}{2}\right)^{2\eta}\frac{1}{(2\eta+1)!} \, \left[\partial_x^{2\eta+1}\mathcal{V}(x)\right]\,\partial_k^{2\eta+1}\mathcal{G}_{\gamma}(x, \, k),
\end{eqnarray}
can thus be worked out in terms of Gaussian relations with Hermite polynomials of order $n$, $\mbox{\sc{h}}_n$,
\begin{equation}
\partial_\chi^{2\eta+1}\mathcal{G}_{\gamma}(x, \, k) = (-1)^{2\eta+1}\gamma^{2\eta+1}\,\mbox{\sc{h}}_{2\eta+1} (\gamma \chi)\, \mathcal{G}_{\gamma}(x, \, k),
\end{equation}
for $\chi = x,\, k$, which can be reintroduced into Eqs.~\eqref{eqnimWA2} and \eqref{eqnimWB2} as to lead to convergent series expansions which account for the overall quantum fluctuations over a classical background.

According to the results from Refs. \cite{2021A,2021B,2021C}, contributions at Eqs.~\eqref{eqnimWA2} and \eqref{eqnimWB2} where $\mathcal{V}$ and $\mathcal{K}$ derivatives can be replaced by
\begin{eqnarray}
\label{eqnt111}
\partial_x^{2\eta+1}\mathcal{V}(x) &=& \lambda^{2\eta+1}_{(x)} \, \upsilon(x),\\
\label{eqnt222}
\partial_k^{2\eta+1}\mathcal{K}(k) &=& \mu^{2\eta+1}_{(k)} \, \kappa(k),
\end{eqnarray}
and substituted into Eqs.~\eqref{eqnimWA2} and \eqref{eqnimWB2}, as to result into
\begin{eqnarray}
\label{eqnimWA3}\partial_x\mathcal{J}_x(x, \, k) &=& (+2i) \kappa(k)\,\mathcal{G}_{\gamma}(x, \, k)\,\sum_{\eta=0}^{\infty} \left(\frac{i\,\gamma\,\mu_{(k)}}{2}\right)^{2\eta+1}\frac{1}{(2\eta+1)!} \, \mbox{\sc{h}}_{2\eta+1} (\gamma k),\\
\label{eqnimWB3}\partial_k\mathcal{J}_k(x, \, k) &=& (-2i) \upsilon(x)\,\mathcal{G}_{\gamma}(x, \, k)\sum_{\eta=0}^{\infty} \left(\frac{i\,\gamma\, \lambda_{(x)}}{2}\right)^{2\eta+1}\frac{1}{(2\eta+1)!} \, \mbox{\sc{h}}_{2\eta+1} (\gamma x),\end{eqnarray}
which can be put in the final form of
\begin{eqnarray}
\label{eqnimWA4}\partial_x\mathcal{J}_x(x, \, k) &=& -2 \kappa(k)\,\sin\left(\gamma^2 \mu_{(k)}\,x\right)\,\exp[+\gamma^2 \mu^2_{(k)}/4]\,\mathcal{G}_{\gamma}(x, \, k)\,
,\\
\label{eqnimWB4}\partial_k\mathcal{J}_k(x, \, k) &=& +2 \upsilon(x)\,\sin\left(\gamma^2 \lambda_{(x)}\,k\right)\,\exp[+\gamma^2 \lambda^2_{(k)}/4]\,\mathcal{G}_{\gamma}(x, \, k),
\end{eqnarray}
where it has been considered that
\begin{equation}
\sum_{\eta=0}^{\infty}\mbox{\sc{h}}_{2\eta+1} (\gamma \chi)\frac{s^{2\eta+1}}{(2\eta+1)!} = \sinh(2s\,\gamma\chi) \exp[-s^2].
\end{equation}

From Eqs.~\eqref{eqnimWA4} and \eqref{eqnimWB4}, one has the convergent series expression for the stationarity quantifier, $\mbox{\boldmath $\nabla$}_{\xi}\cdot \mbox{\boldmath $\mathcal{J}$}$ and, depending on the Hamiltonian dynamics, it can be manipulated to return additional quantum information quantifiers which provide a quantitative view of the associated Wigner flow.

\end{document}